\RequirePackage{lineno}
\documentclass[aps,prl,twocolumn,letterpaper,superscriptaddress, showpacs, amsmath, amssymb]{revtex4}
\usepackage{graphicx}
\usepackage{amsmath,amssymb}
\usepackage{color}

\usepackage{mathrsfs}
\usepackage[x11names]{xcolor}
\usepackage{aas_macros}
\usepackage[normalem]{ulem}

\begin{document}


\title{Hidden Cosmic-Ray Accelerators as an Origin of TeV-PeV Cosmic Neutrinos}
\author{Kohta Murase}
\affiliation{Center for Particle and Gravitational Astrophysics; Department of Physics; Department of Astronomy and Astrophysics, 
The Pennsylvania State University, University Park, Pennsylvania 16802, USA}
\affiliation{Institute for Advanced Study, Princeton, New Jersey 08540, USA}
\author{Dafne Guetta}
\affiliation{Osservatorio Astronomico di Roma, I-00040 Monte Porzio Catone, Italy}
\affiliation{Department of Physics and Optical Engineering, ORT Braude College, Karmiel 21982, Israel} 
\author{Markus Ahlers}
\affiliation{Wisconsin IceCube Particle Astrophysics Center (WIPAC) and Department of Physics, University of Wisconsin, Madison, Wisconsin 53706, USA}

\date{posted 2 September 2015; submitted 7 September 2015; accepted 9 January 2016}

\begin{abstract}
The latest IceCube data suggest that the all-flavor cosmic neutrino flux may be as large as ${10}^{-7}~{\rm GeV}~{\rm cm}^{-2}~{\rm s}^{-1}~{\rm sr}^{-1}$ around 30~TeV.  We show that, if sources of the TeV-PeV neutrinos are transparent to $\gamma$ rays with respect to two-photon annihilation, strong tensions with the isotropic diffuse $\gamma$-ray background measured by {\it Fermi} are unavoidable, independently of the production mechanism. 
We further show that, if the IceCube neutrinos have a photohadronic ($p\gamma$) origin, the sources are expected to be opaque to $1\mbox{--}100$~GeV $\gamma$ rays.  With these general multimessenger arguments, we find that the latest data suggest a population of cosmic-ray accelerators hidden in GeV-TeV $\gamma$ rays as a neutrino origin.  Searches for x-ray and MeV $\gamma$-ray counterparts are encouraged, and TeV-PeV neutrinos themselves will serve as special probes of dense source environments.  
\end{abstract}

\pacs{95.85.Ry, 98.70.Sa, 98.70.Vc\vspace{-0.3cm}}
\maketitle

%
The astrophysical high-energy neutrino flux observed with IceCube~\cite{Aartsen:2013bka,Aartsen:2013jdh,Aartsen:2014gkd,Aartsen:2014muf,Aartsen:2015ita,Aartsen:2015rwa,Botner:2015ipa} is consistent with an isotropic distribution of arrival directions, suggesting a significant contribution from extragalactic neutrino sources.  Most likely, the neutrino signals are generated in the decay of charged pions produced in inelastic hadronuclear ($pp$) and/or photohadronic ($p\gamma$) processes of cosmic rays (CRs)~\cite{Laha:2013lka,Waxman:2013zda,Meszaros:2014tta,Murase:2014tsa}.  All these processes also predict the generation of hadronic $\gamma$ rays from the production and decay of neutral pions.  The power of multimessenger constraints of astrophysical scenarios has been demonstrated~\cite{Murase:2013rfa} in light of the IceCube and {\it Fermi} data~\cite{Ackermann:2014usa}.  CR reservoirs such as starburst galaxies and galaxy clusters or groups have been considered as promising sources, and neutrinos produced by $pp$ interactions between CRs and gas could account for the diffuse flux at $\gtrsim100$~TeV~\cite{Loeb:2006tw,Murase:2008yt,Kotera:2009ms,Murase:2013rfa}.   

The contribution of astrophysical neutrinos has been studied based on various analysis techniques.  By now, the strongest significance comes from high-energy starting event (HESE) searches with IceCube~\cite{Aartsen:2013bka,Aartsen:2013jdh,Botner:2015ipa}.  A recent combined likelihood analysis~\cite{Aartsen:2015ita} sensitive to neutrino energies of 10~TeV to 10~PeV suggests the all-flavor flux is $E_\nu^2\Phi_\nu^{\rm IC}\sim{10}^{-7}~{\rm GeV}~{\rm cm}^{-2}~{\rm s}^{-1}~{\rm sr}^{-1}$ around 30~TeV and a power-law index $s_{\rm ob}=2.50\pm0.09$ (for $\Phi_\nu^{\rm IC}\propto E_\nu^{-s_{\rm ob}}$).  The most recent HESE data also indicate such a soft component~\cite{Botner:2015ipa}.  These observations are consistent with an equal contribution of three neutrino flavors~\cite{Aartsen:2015ivb,Palomares-Ruiz:2015mka,Palladino:2015zua,Bustamante:2015waa}. 

This work considers multimessenger implications of the latest IceCube results for an extragalactic origin.  As shown in Ref.~\cite{Murase:2013rfa}, the neutrino and $\gamma$-ray spectral index should be $s\lesssim2.1\mbox{--}2.2$ for a power-law $\Phi_{\nu/\gamma}\propto E_{\nu/\gamma}^{-s}$, in contrast to $s_{\rm ob}\approx2.4\mbox{--}2.6$.  In CR reservoir models explaining $\lesssim100$~TeV data, the spectral index should be $s\sim2.0$ and $\sim100$\% of the isotropic diffuse $\gamma$-ray background (IGRB) comes from the same neutrino sources, challenging the $pp$ scenarios.  Our results motivate us to study $p\gamma$ scenarios such as models of choked gamma-ray burst (GRB) jets~\cite{Murase:2013ffa} and active galactic nuclei (AGN) cores~\cite{Winter:2013cla,Stecker:2013fxa,Kimura:2014jba}, which are opaque to GeV-TeV $\gamma$ rays. 

\begin{figure*}[bt]
\includegraphics[width=\linewidth]{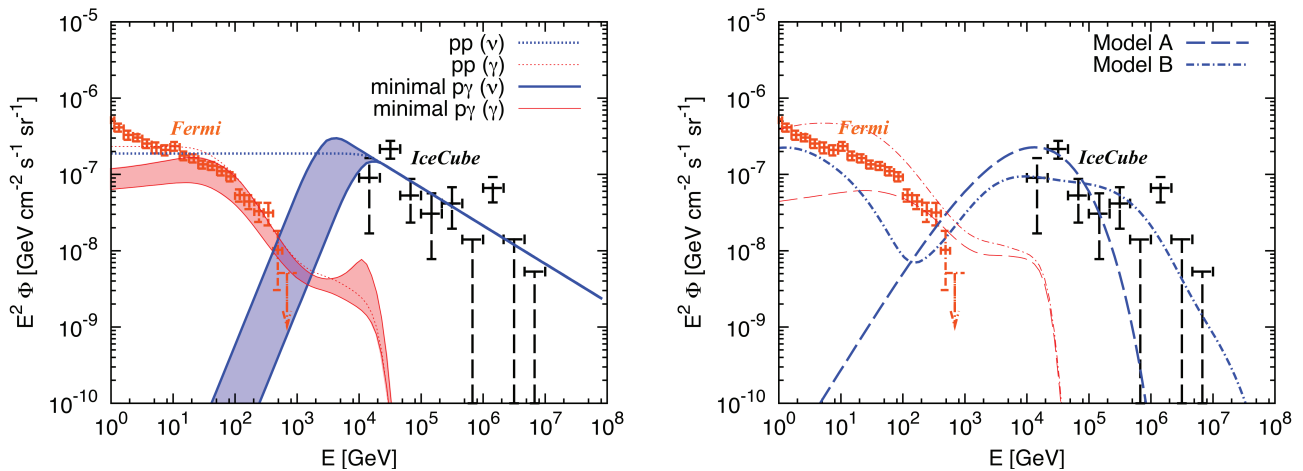}
\caption{{\bf Left panel:} All-flavor neutrino (thick blue lines) and isotropic diffuse $\gamma$-ray (thin red lines) fluxes for $pp$ and {\it minimal} $p\gamma$ scenarios of Eqs.~(\ref{eq:pp}) and (\ref{eq:pgamma}) that account for the latest IceCube data from $\sim10$~TeV to $\sim2$~PeV energies~\cite{Aartsen:2015ita}, where $s'=s_{\rm ob}=2.5$ is used.  While $pp$ scenarios require $\varepsilon_\nu^b=25$~TeV with a strong tension with the {\it Fermi} IGRB~\cite{Ackermann:2014usa}, {\it minimal} $p\gamma$ scenarios allow the range $\varepsilon_\nu^b$ of $6\mbox{--}25$~TeV (shaded regions) as long as the sources are transparent to $\gamma$ rays (see the main text for details).  
{\bf Right panel:} Same as the left panel, but now showing neutrino fluxes of AGN core and choked jet models from Refs.~\cite{Kimura:2014jba,Murase:2013ffa}.  To illustrate the strength of diffuse $\gamma$-ray constraints, we pretend that the sources were transparent to $\gamma$ rays.
\label{fig1}
}
\end{figure*}

{\bf Connecting $\nu$ and $\gamma$ fluxes.---}
Hadronic interactions of CRs lead to the production of mesons (mostly pions), which generates a flux of neutrinos via decay processes like $\pi^+\to\mu^+\nu_\mu$ followed by $\mu^+\to e^+\nu_e\bar{\nu}_\mu$.  The neutrino energy $\varepsilon_\nu$ (in the cosmic reference frame) is related to the proton energy $\varepsilon_p$ as $\varepsilon_\nu\sim(0.04\mbox{--}0.05)\varepsilon_p$. The neutrino energy generation rate $\varepsilon_{\nu}Q_{\varepsilon_\nu}$ is given by 
\begin{equation}\label{eq:Qnu}
\varepsilon_{\nu}Q_{\varepsilon_\nu}\approx\frac{3K}{4(1+K)}{\rm min}[1,f_{pp/p\gamma}]\varepsilon_pQ_{\varepsilon_p}\,,
\end{equation}
where $\varepsilon_{p}Q_{\varepsilon_p}$ is the CR generation rate.  Here the factor $3/4$ accounts for the $1/4$ energy loss for the production of $e^\pm$ in the previous decay chain and $K$ denotes the average ratio of charged to neutral pions with $K\approx1$ for $p\gamma$ and $K\approx2$ for $pp$ interactions.  The energy-dependent meson production efficiency, ${\rm min}[1,f_{pp/p\gamma}]$, accounts for the source environment.  The corresponding all-flavor diffuse neutrino flux, $\Phi_{\nu}$, is calculated as~(e.g., Refs.~\cite{Murase:2012df,Ahlers:2011sd})
\begin{equation}\label{eq:Phinu}
E_{\nu}^2\Phi_{\nu}=\frac{c}{4\pi}\!\int\!\frac{{\rm d}z}{(1+z)^2H(z)}[\varepsilon_{\nu}Q_{\varepsilon_\nu}(z)]\big|_{\varepsilon_\nu=(1+z)E_\nu}\,,
\end{equation}
where $E_\nu$ is the observed neutrino energy and $H(z)$ is the redshift-dependent Hubble parameter. 

The decay of neutral pions $\pi^0\rightarrow2\gamma$ leads to $\gamma$-ray emission.  On production, the neutrino and $\gamma$-ray energy generation rates are conservatively related as~\footnote{Equation~(\ref{eq:Qgamma}) is conservative.  In reality, the generated $\gamma$-ray flux can be higher than the neutrino flux (e.g., Refs.~\cite{Meszaros:2001ms,Murase:2013ffa}), since neutrino emission can be significantly suppressed due to radiative or hadronic or adiabatic cooling of charged pions and muons.  Emission from electrons and positrons that are produced by Bethe-Heitler and photomeson production processes also enhances the relative $\gamma$-ray flux.}
\begin{equation}\label{eq:Qgamma}
\varepsilon_{\gamma}Q_{\varepsilon_\gamma}\approx\frac{4}{3K}(\varepsilon_{\nu}Q_{\varepsilon_\nu})\big|_{\varepsilon_\nu=\varepsilon_\gamma/2}\,,
\end{equation}
where $\gamma$-ray and neutrino energies are related as $\varepsilon_\gamma\approx2\varepsilon_\nu$. However, the {\it generated} $\gamma$ rays from the sources may not be directly observable.  First, $\gamma$ rays above TeV energies initiate electromagnetic cascades in the extragalactic background light (EBL) and cosmic microwave background (CMB) as they propagate over cosmic distances.  As a result, high-energy $\gamma$ rays are regenerated at sub-TeV energies~\cite{Berezinsky:1975zz}.  Second, intrasource cascades via two-photon annihilation, inverse-Compton scattering, and synchrotron radiation processes can prevent direct $\gamma$-ray escape~\cite{Mannheim:1998wp}.  To see their importance, we {\it temporarily} assume that the sources are $\gamma$-ray transparent.  We will see in the following that this hypothesis leads to strong tensions with the IGRB, disfavored by the {\it Fermi} data. 

In $pp$ scenarios, neutrino and generated $\gamma$-ray spectra follow the CR spectrum, assumed to be a power law.  In CR reservoirs such as galaxies and clusters, a spectral break due to CR diffusion is naturally expected~\cite{Loeb:2006tw,Murase:2008yt}. Thus, the neutrino spectrum is approximately given by   
\begin{equation}\label{eq:pp}
\varepsilon_{\nu}Q_{\varepsilon_\nu}\propto
\begin{cases}
\varepsilon_\nu^{2-s}
& (\varepsilon_\nu\leq \varepsilon_\nu^b)\\
\varepsilon_\nu^{2-s'}
& (\varepsilon_\nu^b<\varepsilon_\nu)\\
\end{cases}
\quad\text{($pp$)}\,,
\end{equation}
where $\varepsilon_\nu^b$ is the break energy and the softening of the spectrum, $\delta\equiv s'-s$, is expected from the energy dependence of the diffusion tensor~\footnote{It depends on characteristics of magnetic turbulence, and typical values are $\delta=1/3$ for the Kolmogorov and $\delta=1/2$ for Kraichnan turbulence}.  
In $pp$ scenarios, the corresponding generated $\gamma$-ray spectrum is also a power law $\varepsilon_\gamma^{-s}$ into the sub-TeV region [see Eq.~(\ref{eq:Qgamma})], where it directly contributes to the IGRB~\footnote{Note that typical CR reservoirs are transparent up to $\sim10\mbox{--}100$~TeV energies~\cite{Murase:2012rd,Lacki:2010ue}} and Ref.~\cite{Murase:2013rfa} obtained a limit $s\lesssim2.1\mbox{--}2.2$ for generic $pp$ scenarios that explain the $\gtrsim100$~TeV neutrino data.  The limit is tighter ($s\sim2.0$) if one relaxes this condition by shifting $\varepsilon_\nu^b$ to $\lesssim30$~TeV to account for the lower-energy data~\cite{Senno:2015tra}.   

Motivated by results of Ref.~\cite{Aartsen:2015ita}, we calculate the diffuse neutrino spectrum using Eq.~(\ref{eq:pp}) with $s=2$ and $s' = 2.5$ and the corresponding $\gamma$-ray spectrum using Eq.~(\ref{eq:Qgamma}).  Following Ref.~\cite{Murase:2012df}, we numerically solve Boltzmann equations to calculate intergalactic cascades, including two-photon annihilation, inverse-Compton scattering, and adiabatic losses.  In the left panel of Fig.~\ref{fig1} we show the resulting all-flavor neutrino and $\gamma$-ray fluxes as thick blue and thin red lines, respectively, in comparison to the {\it Fermi} IGRB and IceCube neutrino data~\cite{Aartsen:2015ita}.  To explain the $\lesssim100$~TeV neutrino data, the contribution to the IGRB should be at the level of 100\% in the 3~GeV to 1~TeV range and softer fluxes with $s\gtrsim2.0$ clearly overshoot the data.  As pointed out by Ref.~\cite{Murase:2013rfa}, this argument is {\it conservative}: the total extragalactic $\gamma$-ray background is dominated by a subclass of AGN, blazars~(e.g., Refs.~\cite{Costamante:2013sva,Inoue:2014ona}), and their main emission is typically variable and unlikely to be of $pp$ origin~\cite{Atoyan:2001ey,Murase:2014foa}.  Most of the high-energy IGRB is believed to be accounted for by unresolved blazars~\cite{DiMauro:2014wha,Ackermann:2015gga,Ajello:2015mfa}.  Although the IGRB should be decomposed with caution, if this blazar interpretation is correct, there is little room for CR reservoirs~\cite{Murase:2013rfa}. 

In $p\gamma$ scenarios, neutrino and $\gamma$-ray spectra depend on a target photon spectrum.  The effective optical depth to photomeson production ($f_{p\gamma}$) typically increases with CR energy, so that the neutrino spectrum is harder than the CR spectrum.  However, it cannot be too hard since the decay kinematics of pions gives $\varepsilon_\nu Q_{\varepsilon_\nu}\propto{\varepsilon}_\nu^2$ as a low-energy neutrino spectrum~\cite{Gaisser:1990vg}.  In {\it minimal} $p\gamma$ scenarios, where neutrinos with $\varepsilon_\nu\lesssim\varepsilon_\nu^b\lesssim25$~TeV are produced by CRs at the pion production threshold, the neutrino spectrum is approximately given by
\begin{equation}\label{eq:pgamma}
\varepsilon_{\nu}Q_{\varepsilon_\nu}\propto
\begin{cases}
\varepsilon_\nu^{2}
& (\varepsilon_\nu\leq \varepsilon_\nu^b)\\
\varepsilon_\nu^{2-s'}
& (\varepsilon_\nu^b<\varepsilon_\nu)
\end{cases}\quad\text{(minimal $p\gamma$)}\,.
\end{equation}
In the left panel of Fig.~\ref{fig1}, we show the resulting neutrino and $\gamma$-ray spectra with the diffuse neutrino flux and the IGRB~\footnote{Hadronic multi-TeV $\gamma$ rays escaping from the sources will lead to extended pair-halo or diffuse emission, rather than point-source emission~\cite{Murase:2012df}.  Since the direct hadronic $\gamma$-ray emission in the {\it Fermi} band is negligible for optically thin minimal $p\gamma$ scenarios considered here, it is more reasonable to compare to the IGRB.  Note that main $\gamma$-ray emission from resolved blazars is not such intergalactic cascade emission.  Once we conclude that intrasource cascades should occur, comparisons to the total extragalactic $\gamma$-ray background become reasonable for blazars.} for a neutrino break $\varepsilon_\nu^b$ in the range $6\mbox{--}25$~TeV.  Since the sub-TeV emission is dominated by $\gamma$ rays from cascades in the CMB and EBL, the tension with the IGRB can be weaker than in $pp$ scenarios.  However, the IGRB contribution is still at the level of $\sim50$\% for $\varepsilon_\nu^b=25$~TeV and reaches $\sim100$\% for $\varepsilon_\nu^b=6$~TeV.

The spectrum (\ref{eq:pgamma}) can be realized when the target photon spectrum is a power law with a high-energy cutoff or a gray body (see below).  We note that specific models have larger contributions to the IGRB, by accounting for the detailed energy dependence of $f_{pp/p\gamma}$, the contribution from low-energy CRs, and cooling of charged mesons and muons.  As examples, we consider hadronic $\gamma$ rays in the low-luminosity AGN model of Ref.~\cite{Kimura:2014jba} (model A), which can explain $\lesssim100$~TeV neutrino data, and the choked GRB jet model of Ref.~\cite{Murase:2013ffa} (model B), although these sources are predicted to be {\it opaque} to very-high-energy $\gamma$ rays.  The right panel of Fig.~\ref{fig1} shows the corresponding all-flavor neutrino and generated $\gamma$-ray spectra as thick blue and thin red lines.  Pretending $\gamma$-ray transparency leads to violation of the high-energy IGRB data.  

The limits of the IGRB contribution of $p\gamma$ scenarios are expected to become even stronger by identifying additional point sources or by decomposing the emission into contributions from individual source populations.  This should further constrain the $\gamma$-ray transparent sources for $\varepsilon_\nu^b=6\mbox{--}25$~TeV, which may still be allowed by the {\it Fermi} data (cf. left panel of Fig.~\ref{fig1}). 
On the other hand, since the sub-TeV $\gamma$-ray emission comes from cascades in the CMB and EBL, the tension with the IGRB can easily be relaxed compared to $pp$ scenarios if the sources are {\it $\gamma$-ray dark}, {\it i.e.} if high-energy $\gamma$ rays generated in the sources of diffuse neutrinos undergo efficient interactions with intrasource radiation. 
In fact, this is generally the case for $p\gamma$ scenarios as we will show in the following.

{\bf Connecting $p\gamma$ and $\gamma\gamma$ optical depths.---}
Let us consider a generic source with comoving size $r/\Gamma$ (where $r$ is the emission radius and $\Gamma$ is the bulk Lorentz factor of the source).  We assume the presence of target photons with $\varepsilon'_t\approx\varepsilon_t/\Gamma$ and spectrum $n_{\varepsilon'_t}$.  
For $n_{\varepsilon'_t}\propto{\varepsilon'}_t^{-\alpha}$ with $\alpha>1$, which is valid in most nonthermal objects, meson production is dominated by the $\Delta$-resonance and direct pion production.  Its efficiency $f_{p\gamma}$ is given by
\begin{equation}\label{eq:fpgamma}
f_{p \gamma}(\varepsilon_p)\approx\eta_{p\gamma}(\alpha)\hat{\sigma}_{p\gamma}(r/\Gamma)(\varepsilon'_t n_{\varepsilon'_t})|_{\varepsilon'_t=0.5m_pc^2\bar{\varepsilon}_{\Delta}/\varepsilon'_p}\,,
\end{equation}
where $\hat{\sigma}_{p\gamma}\sim0.7\times{10}^{-28}~{\rm cm}^2$ is the attenuation cross section (the product of the inelasticity and cross section~\cite{Dermer:2010hy,Murase:2008mr,Dermer:2014vaa}), $\eta_{p\gamma}(\alpha)\approx2/(1+\alpha)$, and $\bar{\varepsilon}_\Delta\sim0.3$~GeV.  The energy of protons that typically interact with photons with cosmic reference frame energy $\varepsilon_t$ is $\varepsilon_p\approx20\varepsilon_\nu\approx0.5\Gamma^2m_pc^2\bar{\varepsilon}_\Delta{\varepsilon_t}^{-1}$, leading to $\varepsilon_t\sim20~{\rm keV}~{(\Gamma/10)}^2{(\varepsilon_\nu/{\rm 30~TeV})}^{-1}$.  Thus, the IceCube data imply sources with {\it x-ray or MeV $\gamma$-ray counterparts}.  
If target radiation is generated by synchrotron or inverse-Compton emission from thermal or nonthermal electrons, low-energy photon spectra can be expressed by power-law segments, $n_{\varepsilon'_t}\propto{\varepsilon'}_t^{-\alpha}$, where $\alpha\geq2/3$~\cite{Dermer:2010hy}.  For $n_{\varepsilon'_p}\propto{\varepsilon'}_p^{-s_{\rm cr}}$ and $\alpha\gtrsim1$, the efficiency scales as $f_{p\gamma}\propto\varepsilon_p^{\alpha-1}$, and the neutrino spectral index is $s=s_{\rm cr}+1-\alpha$. 
For $\alpha\lesssim1$ the secondary neutrino and $\gamma$-ray spectra follow the initial CR spectrum, $s\sim s_{\rm cr}$, above the pion production threshold because $f_{p\gamma}$ becomes approximately constant due to higher resonances and multipion production~\cite{Dermer:2014vaa,Murase:2008mr}.  A similar scaling is obtained for gray-body and monochromatic target photon spectra~\cite{Murase:2014foa,Dermer:2014vaa}.  

Now, in $p\gamma$ scenarios, the same target photon field can prevent $\gamma$ rays from escaping the sources.  The relevance of two-photon annihilation in GRBs and AGN has been considered~\cite{Rees:1967,Rees:1992ek}.  The optical depth to $\gamma\gamma\rightarrow e^+e^-$ is
\begin{equation}\label{eq:taugammagamma}
\tau_{\gamma\gamma}(\varepsilon_\gamma)\approx\eta_{\gamma\gamma}(\alpha)\sigma_{T}(r/\Gamma)(\varepsilon'_t n_{\varepsilon'_t})|_{\varepsilon'_t=m_e^2c^4/\varepsilon'_\gamma}\,,
\end{equation}
where $\sigma_T\simeq6.65\times{10}^{-25}~{\rm cm}^2$ and $\eta_{\gamma\gamma}(\alpha)\simeq7/[6\alpha^{5/3}(1+\alpha)]$ for $1<\alpha<7$~\cite{sve87}, which is the order of $0.1$.  The typical $\gamma$-ray energy is given by $\varepsilon_\gamma\approx\Gamma^2m_e^2c^4{\varepsilon_t}^{-1}$.

Neutrino sources considered here include transrelativistic or relativistic sources like GRBs, pulsars, and AGN including blazars.  For example, the observed neutrino energy is expressed to be $E_\nu=\varepsilon_\nu/(1+z)\approx\Gamma{\varepsilon'}_\nu/(1+z)$.  Equations~(\ref{eq:fpgamma}) and (\ref{eq:taugammagamma}) can be used for both internal and external photon fields.  As shown in Refs.~\cite{Murase:2014foa,Dermer:2014vaa} for reprocessed radiation from ionized clouds, the cases of $\Gamma=1$ are reduced to the formulas for external photon fields.  Thus, regardless of these model details, Eqs.~(\ref{eq:fpgamma}) and (\ref{eq:taugammagamma}) lead to the following relation~\cite{Murase:2008mr,Dermer:2007me,Mannheim:1998wp}:
\begin{equation}\label{eq:opticaldepth}
\tau_{\gamma\gamma}(\varepsilon_\gamma^{c})\approx\frac{\eta_{\gamma\gamma}\sigma_{\gamma\gamma}}{\eta_{p\gamma}\hat{\sigma}_{p\gamma}}f_{p\gamma}(\varepsilon_p)\sim{10}\left(\frac{f_{p\gamma}(\varepsilon_p)}{0.01}\right)\,,
\end{equation}
where $\varepsilon_\gamma^{c}$ is the $\gamma$-ray energy corresponding to the resonance proton energy satisfying
\begin{equation}\label{eq:gammanurel}
\varepsilon_\gamma^c\approx\frac{2m_e^2c^2}{m_p\bar{\varepsilon}_\Delta}\varepsilon_p\sim{\rm GeV}~\left(\frac{\varepsilon_\nu}{25~{\rm TeV}}\right)\,.
\end{equation}
Thus, the neutrino data from $25$~TeV to $2.8$~PeV~\cite{Aartsen:2015ita}, corresponding to the proton energy range from $\sim0.5$ to $\sim60$~PeV, can directly constrain the two-photon annihilation optical depth at $\varepsilon_\gamma\sim1\mbox{--}100$~GeV.  
Importantly, Eqs.~(\ref{eq:opticaldepth}) and (\ref{eq:gammanurel}) are independent of $\Gamma$ and valid for both internal and external radiation fields. 

\begin{figure}[bt]
\includegraphics[width=\linewidth]{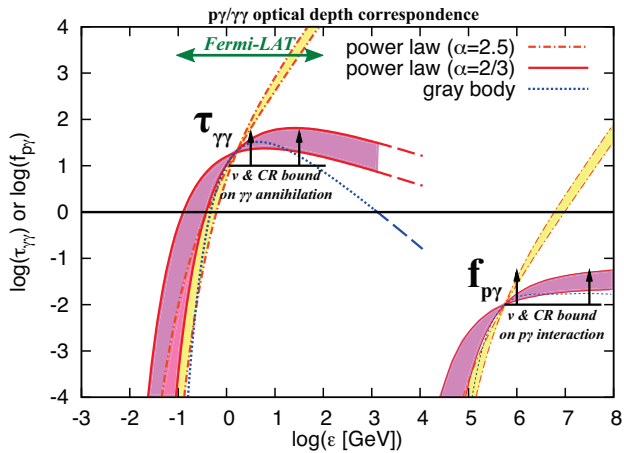}
\caption{Neutrino and CR bounds on the optical depth to $\gamma\gamma\rightarrow e^+e^-$ in the sources of diffuse TeV-PeV neutrinos.  We calculate $\tau_{\gamma\gamma}$ and $f_{p\gamma}$ as functions of $\varepsilon_\gamma$ and $\varepsilon_p$, respectively, imposing $f_{p\gamma}\geq0.01$.  We consider simple power laws with $\alpha=2.5$ and $\alpha=2/3$ for $\varepsilon_\nu^b=6\mbox{--}25$~TeV (shaded bands), and the gray-body case with the temperature $kT/\Gamma^2=112$~eV.  
\label{fig2}
}
\end{figure}

In general, the effective $p\gamma$ optical depth $f_{p\gamma}$ depends on source models.  But too small values of $f_{p\gamma}$ seem unnatural since the observed neutrino flux is not far from the Waxman-Bahcall~\cite{Waxman:1998yy,Bahcall:1999yr} (see also Ref.~\cite{Mannheim:1998wp}) and nucleus-survival bounds~\cite{Murase:2010gj}, corresponding to maximally efficient neutrino production in the sources of ultrahigh-energy (UHE) CRs.  More quantitatively, it is possible to obtain general constraints on $f_{p\gamma}$ by comparing the observed CR and neutrino fluxes. 
Recently, Ref.~\cite{Yoshida:2014uka} obtained $f_{p\gamma}\gtrsim0.01$ by requiring that the extragalactic CR flux does not overshoot the observed all-particle CR flux $E_{\rm cr}^2\Phi_{\rm cr}\approx4\times{10}^{-5}~{\rm GeV}~{\rm cm}^{-2}~{\rm s}^{-1}~{\rm sr}^{-1}$ at $10$~PeV (e.g., Ref.~\cite{Gaisser:2013bla}).  Since the observed CR flux in this energy range is dominated by heavy nuclei from Galactic sources such as supernova remnants, this constraint is conservative.  The recent KASCADE-Grande data~\cite{Apel:2013ura} suggest that a light CR component may become prominent above the second knee energy at 100 PeV, which can be interpreted as the onset of an extragalactic component.  Using their inferred extragalactic, light CR flux $E_p^2\Phi_p\approx2\times{10}^{-6}~{\rm GeV}~{\rm cm}^{-2}~{\rm s}^{-1}~{\rm sr}^{-1}$ as an upper limit, we obtain $f_{p\gamma}\gtrsim0.1$ at $\varepsilon_p\gtrsim10$~PeV~\footnote{These constraints can be relaxed if CRs are largely confined in the source environment, as expected in some CR reservoir models.}.  
 
A similar conclusion is drawn by examining nonthermal luminosity densities of known objects.  The CR luminosity density of galaxies including starbursts is restricted as $\varepsilon_pQ_{\varepsilon_p}\lesssim{10}^{45}\mbox{--}{10}^{46}~{\rm erg}~{\rm Mpc}^{-3}~{\rm yr}^{-1}$~\cite{Katz:2013ooa,Lacki:2013ata}.  The luminosity density of x rays ($Q_X\approx2\times{10}^{46}~{\rm erg}~{\rm Mpc}^{-3}~{\rm yr}^{-1}$~\cite{Ueda:2014tma}), which are thought to originate from thermal electrons in hot coronae, can be regarded as an upper limit of nonthermal outputs from AGN.  Adopting $\varepsilon_pQ_{\varepsilon_p}\lesssim2\times{10}^{46}~{\rm erg}~{\rm Mpc}^{-3}~{\rm yr}^{-1}$ as a reasonable assumption for CRs from galaxies or AGN, we have $f_{p\gamma}\gtrsim0.01$, independently of the above argument.

Figure~\ref{fig2} shows comparisons of the effective $p\gamma$ optical depth required from the IceCube observation to the corresponding optical depth to $\gamma\gamma$ interactions in the {\it Fermi} range, related by Eq.~(\ref{eq:opticaldepth}).  Strictly speaking, Eqs.~(\ref{eq:opticaldepth}) and (\ref{eq:gammanurel}) are valid for soft target spectra.  To see the robustness of our results, following Ref.~\cite{Murase:2008mr}, we perform numerical calculations using the detailed cross sections of the two-photon annihilation and photomeson production (including nonresonant processes).  We consider target photon spectra leading to $\varepsilon_\nu^b=6\mbox{--}25$~TeV (indicated as bands in Fig.~\ref{fig2}), which can reproduce minimal $p\gamma$ scenarios.  Note that adopting lower values of $\varepsilon_\nu^b$ or assuming $\gamma$-ray transparency for models like those shown in the right panel of Fig.~\ref{fig1} leads to inconsistency with the {\it Fermi} IGRB data.  The conclusion from Eq.~(\ref{eq:opticaldepth}) holds even for realistic target radiation fields, including synchrotron and gray-body spectra. 

The high $p\gamma$ efficiency suggested by the IceCube data and upper limits on CR luminosity densities suggest that the direct $1\mbox{--}100$~GeV $\gamma$-ray emission from the sources--either leptonic or hadronic--is suppressed.  Thus, tensions with the IGRB, which are unavoidable for $\gamma$-ray transparent sources, are largely alleviated or even absent.  However, TeV $\gamma$-ray counterparts could be seen by Cherenkov telescopes and the High-Altitude Water Cherenkov Observatory.  For power-law target photon spectra, which extend to low energies, $\tau_{\gamma\gamma}$ is larger than unity beyond the {\it Fermi} band and as a result the TeV emission from the sources should also be suppressed (see Fig.~\ref{fig2}).  For gray-body-like spectra, one could expect point-source $\gamma$-ray emission above TeV.  The escaping hadronic $\gamma$ rays are cascaded in the CMB and EBL and could be visible as extended pair-halo emission in the sub-TeV range (e.g., Refs.~\cite{Murase:2012df,Ahlers:2011sd}).  In this special case, although direct point-source emission at $1\mbox{--}100$~GeV is still suppressed and the tension with the IGRB remains, TeV counterpart searches can be used as an additional test.  

{\bf Summary and implications.---}
We considered implications of the latest IceCube results in light of the multimessenger data.  Based on the diffuse $\nu$-$\gamma$ flux connection and CR-$\gamma$ optical depth connection, we showed that the two-photon annihilation optical depth should be large as a direct consequence of astrophysical scenarios that explain the large flux observed in IceCube. 
 
There are various implications.  Cross correlation of neutrinos with {\it Fermi-LAT} sources is predicted to be weak.  Rather, in $p\gamma$ scenarios, since target photons are expected in the x-ray or MeV $\gamma$-ray range, searches for such counterparts are encouraged.  
Candidate sources of hidden CR accelerators include choked GRB jets~\cite{Murase:2013ffa} and supermassive black hole cores~\cite{Stecker:2013fxa,Kimura:2014jba,Kalashev:2015cma} (see also the Supplementary Material~\footnote{See Supplementary Material for a summary of $\gamma$-ray dark extragalactic candidate sources and multi-messenger limits on Galactic neutrino sources, which includes Refs.~\cite{Tamborra:2014xia,Meszaros:2015zka,Waxman:1995vg,Vietri:1995hs,Aartsen:2014aqy,Murase:2006mm,Gupta:2006jm,Bhattacharya:2014sta,Nakar:2015tma,Senno:2015tsn,Murase:2009pg,Fang:2013vla,Liang:2006ci,Smith:2010vz,Matsumoto:2015bga,Iocco:2007td,Ishihara:2015tev,Murase:2015ndr,Inoue:2008ks,Ajello:2013lka,Tavecchio:2014eia,Blandford:1979za,Padovani:2015mba,Tavecchio:2014xha,Glusenkamp:2015jca,Stecker:1991vm,AlvarezMuniz:2004uz,Pe'er:2009rc,Burrows:2011dn,Aleksic:2014xsg,Schatz:2003aw,Chantell:1997gs,Borione:1997fy,Ahlers:2015moa,Chen:2014gxa}.}, which includes Refs.~\cite{Tamborra:2014xia,Meszaros:2015zka,Waxman:1995vg,Vietri:1995hs,Aartsen:2014aqy,Murase:2006mm,Gupta:2006jm,Bhattacharya:2014sta,Nakar:2015tma,Senno:2015tsn,Murase:2009pg,Fang:2013vla,Liang:2006ci,Smith:2010vz,Matsumoto:2015bga,Iocco:2007td,Ishihara:2015tev,Murase:2015ndr,Inoue:2008ks,Ajello:2013lka,Tavecchio:2014eia,Blandford:1979za,Padovani:2015mba,Tavecchio:2014xha,Glusenkamp:2015jca,Stecker:1991vm,AlvarezMuniz:2004uz,Pe'er:2009rc,Burrows:2011dn,Aleksic:2014xsg,Schatz:2003aw,Chantell:1997gs,Borione:1997fy,Ahlers:2015moa,Chen:2014gxa}), so correlations with energetic supernovae including low-power GRBs, flares from supermassive black holes, radio-quiet or low-luminosity AGN, and a subclass of flat spectrum radio quasars can be used to test the models. 
For broadband nonthermal target photon spectra, $\gamma$ rays are suppressed at TeV-PeV as well as $1\mbox{--}100$~GeV energies. However, if the target photons follow a narrow thermal spectrum or are monochromatic in x rays, hadronic $\gamma$ rays might be seen in the TeV range for nearby neutrino sources.  Although the obvious multimessenger relation between neutrinos and $\gamma$ rays no longer exists, our findings suggest that cosmic neutrinos play a special role in the study of dense source environments that are not probed by $\gamma$ rays.  Larger detectors such as {\it IceCube-Gen2}~\cite{Aartsen:2014njl} sensitive to $10\mbox{--}100$~TeV neutrinos would be important for the identification of the sources via autocorrelation of neutrino events~\cite{Ahlers:2014ioa,mw15}.

We have assumed that the diffuse neutrino emission is isotropic.  
Even if half of the neutrino flux has a Galactic origin, which allows somewhat smaller values of $\varepsilon_\nu^b\sim2$~TeV and $f_{p\gamma}$, our conclusions remain unchanged.  
Future data on the arrival distribution of starting muon events will also be useful. 

\medskip
\begin{acknowledgments}
We thank Markus Ackermann, John Beacom, Francis Halzen, and Shigeru Yoshida for discussions.  
K.~M. acknowledges the Institute for Advanced Study and Penn State University for support.  K.~M. also thanks the INT Program ``Neutrino Astrophysics and Fundamental Properties'' during the development of part of this work.  M.~A. acknowledges support by the U.S. National Science Foundation (NSF) under Grants No. OPP-0236449 and No. PHY-0236449.  D.~G. is supported by a grant from the U.S. Israel Binational Science Foundation.  

{\it Note added.---}
After this work came out and was submitted, Refs.~\cite{Ando:2015bva,Bechtol:2015uqb} have recently appeared and support our argument on $pp$ scenarios. 
\end{acknowledgments}


\bibliography{kmurase.bib}

\hyphenation{Post-Script Sprin-ger}
\begin{thebibliography}{97}
\expandafter\ifx\csname natexlab\endcsname\relax\def\natexlab#1{#1}\fi
\expandafter\ifx\csname bibnamefont\endcsname\relax
  \def\bibnamefont#1{#1}\fi
\expandafter\ifx\csname bibfnamefont\endcsname\relax
  \def\bibfnamefont#1{#1}\fi
\expandafter\ifx\csname citenamefont\endcsname\relax
  \def\citenamefont#1{#1}\fi
\expandafter\ifx\csname url\endcsname\relax
  \def\url#1{\texttt{#1}}\fi
\expandafter\ifx\csname urlprefix\endcsname\relax\def\urlprefix{URL }\fi
\providecommand{\bibinfo}[2]{#2}
\providecommand{\eprint}[2][]{\url{#2}}

\bibitem[{\citenamefont{Aartsen et~al.}(2013{\natexlab{a}})}]{Aartsen:2013bka}
\bibinfo{author}{\bibfnamefont{M.}~\bibnamefont{Aartsen}} \bibnamefont{et~al.}
  (\bibinfo{collaboration}{IceCube Collaboration}),
  \bibinfo{journal}{Phys.Rev.Lett.} \textbf{\bibinfo{volume}{111}},
  \bibinfo{pages}{021103} (\bibinfo{year}{2013}{\natexlab{a}}),
  \eprint{1304.5356}.

\bibitem[{\citenamefont{Aartsen et~al.}(2013{\natexlab{b}})}]{Aartsen:2013jdh}
\bibinfo{author}{\bibfnamefont{M.}~\bibnamefont{Aartsen}} \bibnamefont{et~al.}
  (\bibinfo{collaboration}{IceCube Collaboration}), \bibinfo{journal}{Science}
  \textbf{\bibinfo{volume}{342}}, \bibinfo{pages}{1242856}
  (\bibinfo{year}{2013}{\natexlab{b}}), \eprint{1311.5238}.

\bibitem[{\citenamefont{Aartsen et~al.}(2014{\natexlab{a}})}]{Aartsen:2014gkd}
\bibinfo{author}{\bibfnamefont{M.}~\bibnamefont{Aartsen}} \bibnamefont{et~al.}
  (\bibinfo{collaboration}{IceCube Collaboration}),
  \bibinfo{journal}{Phys.Rev.Lett.} \textbf{\bibinfo{volume}{113}},
  \bibinfo{pages}{101101} (\bibinfo{year}{2014}{\natexlab{a}}),
  \eprint{1405.5303}.

\bibitem[{\citenamefont{Aartsen et~al.}(2015{\natexlab{a}})}]{Aartsen:2014muf}
\bibinfo{author}{\bibfnamefont{M.}~\bibnamefont{Aartsen}} \bibnamefont{et~al.}
  (\bibinfo{collaboration}{IceCube Collaboration}),
  \bibinfo{journal}{Phys.Rev.} \textbf{\bibinfo{volume}{D91}},
  \bibinfo{pages}{022001} (\bibinfo{year}{2015}{\natexlab{a}}),
  \eprint{1410.1749}.

\bibitem[{\citenamefont{Aartsen et~al.}(2015{\natexlab{b}})}]{Aartsen:2015ita}
\bibinfo{author}{\bibfnamefont{M.~G.} \bibnamefont{Aartsen}}
  \bibnamefont{et~al.} (\bibinfo{collaboration}{IceCube Collaboration}),
  \bibinfo{journal}{Astrophys. J.} \textbf{\bibinfo{volume}{809}},
  \bibinfo{pages}{98} (\bibinfo{year}{2015}{\natexlab{b}}),
  \eprint{1507.03991}.

\bibitem[{\citenamefont{Aartsen et~al.}(2015{\natexlab{c}})}]{Aartsen:2015rwa}
\bibinfo{author}{\bibfnamefont{M.~G.} \bibnamefont{Aartsen}}
  \bibnamefont{et~al.} (\bibinfo{collaboration}{IceCube Collaboration}),
  \bibinfo{journal}{Phys. Rev. Lett.} \textbf{\bibinfo{volume}{115}},
  \bibinfo{pages}{081102} (\bibinfo{year}{2015}{\natexlab{c}}),
  \eprint{1507.04005}.

\bibitem[{\citenamefont{Botner}(2015)}]{Botner:2015ipa}
\bibinfo{author}{\bibfnamefont{O.}~\bibnamefont{Botner}}
  (\bibinfo{collaboration}{IceCube Collaboration}), \bibinfo{journal}{in talks
  presented at the IPA Symposium 2015}  (\bibinfo{year}{2015}).

\bibitem[{\citenamefont{Laha et~al.}(2013)\citenamefont{Laha, Beacom, Dasgupta,
  Horiuchi, and Murase}}]{Laha:2013lka}
\bibinfo{author}{\bibfnamefont{R.}~\bibnamefont{Laha}},
  \bibinfo{author}{\bibfnamefont{J.~F.} \bibnamefont{Beacom}},
  \bibinfo{author}{\bibfnamefont{B.}~\bibnamefont{Dasgupta}},
  \bibinfo{author}{\bibfnamefont{S.}~\bibnamefont{Horiuchi}}, \bibnamefont{and}
  \bibinfo{author}{\bibfnamefont{K.}~\bibnamefont{Murase}},
  \bibinfo{journal}{Phys.Rev.} \textbf{\bibinfo{volume}{D88}},
  \bibinfo{pages}{043009} (\bibinfo{year}{2013}), \eprint{1306.2309}.

\bibitem[{\citenamefont{Waxman}(2013)}]{Waxman:2013zda}
\bibinfo{author}{\bibfnamefont{E.}~\bibnamefont{Waxman}}
  (\bibinfo{year}{2013}), \eprint{1312.0558}.

\bibitem[{\citenamefont{M\'esz\'aros}(2014)}]{Meszaros:2014tta}
\bibinfo{author}{\bibfnamefont{P.}~\bibnamefont{M\'esz\'aros}},
  \bibinfo{journal}{Nucl. Phys. Proc. Suppl.}
  \textbf{\bibinfo{volume}{256-257}}, \bibinfo{pages}{241}
  (\bibinfo{year}{2014}), \eprint{1407.5671}.

\bibitem[{\citenamefont{Murase}(2015{\natexlab{a}})}]{Murase:2014tsa}
\bibinfo{author}{\bibfnamefont{K.}~\bibnamefont{Murase}}, \bibinfo{journal}{AIP
  Conf. Proc.} \textbf{\bibinfo{volume}{1666}}, \bibinfo{pages}{040006}
  (\bibinfo{year}{2015}{\natexlab{a}}), \eprint{1410.3680}.

\bibitem[{\citenamefont{Murase et~al.}(2013)\citenamefont{Murase, Ahlers, and
  Lacki}}]{Murase:2013rfa}
\bibinfo{author}{\bibfnamefont{K.}~\bibnamefont{Murase}},
  \bibinfo{author}{\bibfnamefont{M.}~\bibnamefont{Ahlers}}, \bibnamefont{and}
  \bibinfo{author}{\bibfnamefont{B.~C.} \bibnamefont{Lacki}},
  \bibinfo{journal}{Phys.Rev.} \textbf{\bibinfo{volume}{D88}},
  \bibinfo{pages}{121301} (\bibinfo{year}{2013}), \eprint{1306.3417}.

\bibitem[{\citenamefont{Ackermann
  et~al.}(2015{\natexlab{a}})}]{Ackermann:2014usa}
\bibinfo{author}{\bibfnamefont{M.}~\bibnamefont{Ackermann}}
  \bibnamefont{et~al.} (\bibinfo{collaboration}{Fermi LAT Collaboration}),
  \bibinfo{journal}{Astrophys.J.} \textbf{\bibinfo{volume}{799}},
  \bibinfo{pages}{86} (\bibinfo{year}{2015}{\natexlab{a}}), \eprint{1410.3696}.

\bibitem[{\citenamefont{Loeb and Waxman}(2006)}]{Loeb:2006tw}
\bibinfo{author}{\bibfnamefont{A.}~\bibnamefont{Loeb}} \bibnamefont{and}
  \bibinfo{author}{\bibfnamefont{E.}~\bibnamefont{Waxman}},
  \bibinfo{journal}{J. Cosmol. Astropart. Phys.}
  \textbf{\bibinfo{volume}{0605}}, \bibinfo{pages}{003} (\bibinfo{year}{2006}),
  \eprint{astro-ph/0601695}.

\bibitem[{\citenamefont{Murase et~al.}(2008{\natexlab{a}})\citenamefont{Murase,
  Inoue, and Nagataki}}]{Murase:2008yt}
\bibinfo{author}{\bibfnamefont{K.}~\bibnamefont{Murase}},
  \bibinfo{author}{\bibfnamefont{S.}~\bibnamefont{Inoue}}, \bibnamefont{and}
  \bibinfo{author}{\bibfnamefont{S.}~\bibnamefont{Nagataki}},
  \bibinfo{journal}{Astrophys.J.} \textbf{\bibinfo{volume}{689}},
  \bibinfo{pages}{L105} (\bibinfo{year}{2008}{\natexlab{a}}),
  \eprint{0805.0104}.

\bibitem[{\citenamefont{Kotera et~al.}(2009)\citenamefont{Kotera, Allard,
  Murase, Aoi, Dubois et~al.}}]{Kotera:2009ms}
\bibinfo{author}{\bibfnamefont{K.}~\bibnamefont{Kotera}},
  \bibinfo{author}{\bibfnamefont{D.}~\bibnamefont{Allard}},
  \bibinfo{author}{\bibfnamefont{K.}~\bibnamefont{Murase}},
  \bibinfo{author}{\bibfnamefont{J.}~\bibnamefont{Aoi}},
  \bibinfo{author}{\bibfnamefont{Y.}~\bibnamefont{Dubois}},
  \bibnamefont{et~al.}, \bibinfo{journal}{Astrophys.J.}
  \textbf{\bibinfo{volume}{707}}, \bibinfo{pages}{370} (\bibinfo{year}{2009}),
  \eprint{0907.2433}.

\bibitem[{\citenamefont{Aartsen et~al.}(2015{\natexlab{d}})}]{Aartsen:2015ivb}
\bibinfo{author}{\bibfnamefont{M.}~\bibnamefont{Aartsen}} \bibnamefont{et~al.}
  (\bibinfo{collaboration}{IceCube Collaboration}),
  \bibinfo{journal}{Phys.Rev.Lett.} \textbf{\bibinfo{volume}{114}},
  \bibinfo{pages}{171102} (\bibinfo{year}{2015}{\natexlab{d}}),
  \eprint{1502.03376}.

\bibitem[{\citenamefont{Palomares-Ruiz
  et~al.}(2015)\citenamefont{Palomares-Ruiz, Vincent, and
  Mena}}]{Palomares-Ruiz:2015mka}
\bibinfo{author}{\bibfnamefont{S.}~\bibnamefont{Palomares-Ruiz}},
  \bibinfo{author}{\bibfnamefont{A.~C.} \bibnamefont{Vincent}},
  \bibnamefont{and} \bibinfo{author}{\bibfnamefont{O.}~\bibnamefont{Mena}},
  \bibinfo{journal}{Phys.Rev.} \textbf{\bibinfo{volume}{D91}},
  \bibinfo{pages}{103008} (\bibinfo{year}{2015}), \eprint{1502.02649}.

\bibitem[{\citenamefont{Palladino et~al.}(2015)\citenamefont{Palladino,
  Pagliaroli, Villante, and Vissani}}]{Palladino:2015zua}
\bibinfo{author}{\bibfnamefont{A.}~\bibnamefont{Palladino}},
  \bibinfo{author}{\bibfnamefont{G.}~\bibnamefont{Pagliaroli}},
  \bibinfo{author}{\bibfnamefont{F.~L.} \bibnamefont{Villante}},
  \bibnamefont{and} \bibinfo{author}{\bibfnamefont{F.}~\bibnamefont{Vissani}},
  \bibinfo{journal}{Phys.Rev.Lett.} \textbf{\bibinfo{volume}{114}},
  \bibinfo{pages}{171101} (\bibinfo{year}{2015}), \eprint{1502.02923}.

\bibitem[{\citenamefont{Bustamante et~al.}(2015)\citenamefont{Bustamante,
  Beacom, and Winter}}]{Bustamante:2015waa}
\bibinfo{author}{\bibfnamefont{M.}~\bibnamefont{Bustamante}},
  \bibinfo{author}{\bibfnamefont{J.~F.} \bibnamefont{Beacom}},
  \bibnamefont{and} \bibinfo{author}{\bibfnamefont{W.}~\bibnamefont{Winter}},
  \bibinfo{journal}{Phys. Rev. Lett.} \textbf{\bibinfo{volume}{115}},
  \bibinfo{pages}{161302} (\bibinfo{year}{2015}), \eprint{1506.02645}.

\bibitem[{\citenamefont{Murase and Ioka}(2013)}]{Murase:2013ffa}
\bibinfo{author}{\bibfnamefont{K.}~\bibnamefont{Murase}} \bibnamefont{and}
  \bibinfo{author}{\bibfnamefont{K.}~\bibnamefont{Ioka}},
  \bibinfo{journal}{Phys.Rev.Lett.} \textbf{\bibinfo{volume}{111}},
  \bibinfo{pages}{121102} (\bibinfo{year}{2013}), \eprint{1306.2274}.

\bibitem[{\citenamefont{Winter}(2013)}]{Winter:2013cla}
\bibinfo{author}{\bibfnamefont{W.}~\bibnamefont{Winter}},
  \bibinfo{journal}{Phys.Rev.} \textbf{\bibinfo{volume}{D88}},
  \bibinfo{pages}{083007} (\bibinfo{year}{2013}), \eprint{1307.2793}.

\bibitem[{\citenamefont{Stecker}(2013)}]{Stecker:2013fxa}
\bibinfo{author}{\bibfnamefont{F.~W.} \bibnamefont{Stecker}},
  \bibinfo{journal}{Phys.Rev.} \textbf{\bibinfo{volume}{D88}},
  \bibinfo{pages}{047301} (\bibinfo{year}{2013}), \eprint{1305.7404}.

\bibitem[{\citenamefont{Kimura et~al.}(2015)\citenamefont{Kimura, Murase, and
  Toma}}]{Kimura:2014jba}
\bibinfo{author}{\bibfnamefont{S.~S.} \bibnamefont{Kimura}},
  \bibinfo{author}{\bibfnamefont{K.}~\bibnamefont{Murase}}, \bibnamefont{and}
  \bibinfo{author}{\bibfnamefont{K.}~\bibnamefont{Toma}},
  \bibinfo{journal}{Astrophys.J.} \textbf{\bibinfo{volume}{806}},
  \bibinfo{pages}{159} (\bibinfo{year}{2015}), \eprint{1411.3588}.

\bibitem[{\citenamefont{Murase et~al.}(2012)\citenamefont{Murase, Beacom, and
  Takami}}]{Murase:2012df}
\bibinfo{author}{\bibfnamefont{K.}~\bibnamefont{Murase}},
  \bibinfo{author}{\bibfnamefont{J.~F.} \bibnamefont{Beacom}},
  \bibnamefont{and} \bibinfo{author}{\bibfnamefont{H.}~\bibnamefont{Takami}},
  \bibinfo{journal}{J. Cosmol. Astropart. Phys.}
  \textbf{\bibinfo{volume}{1208}}, \bibinfo{pages}{030} (\bibinfo{year}{2012}),
  \eprint{1205.5755}.

\bibitem[{\citenamefont{Ahlers and Salvado}(2011)}]{Ahlers:2011sd}
\bibinfo{author}{\bibfnamefont{M.}~\bibnamefont{Ahlers}} \bibnamefont{and}
  \bibinfo{author}{\bibfnamefont{J.}~\bibnamefont{Salvado}},
  \bibinfo{journal}{Phys. Rev.} \textbf{\bibinfo{volume}{D84}},
  \bibinfo{pages}{085019} (\bibinfo{year}{2011}), \eprint{1105.5113}.

\bibitem[{\citenamefont{Berezinsky and Smirnov}(1975)}]{Berezinsky:1975zz}
\bibinfo{author}{\bibfnamefont{V.~S.} \bibnamefont{Berezinsky}}
  \bibnamefont{and} \bibinfo{author}{\bibfnamefont{A.~{\relax Yu}.}
  \bibnamefont{Smirnov}}, \bibinfo{journal}{Astrophys. Space Sci.}
  \textbf{\bibinfo{volume}{32}}, \bibinfo{pages}{461} (\bibinfo{year}{1975}).

\bibitem[{\citenamefont{Mannheim et~al.}(2000)\citenamefont{Mannheim,
  Protheroe, and Rachen}}]{Mannheim:1998wp}
\bibinfo{author}{\bibfnamefont{K.}~\bibnamefont{Mannheim}},
  \bibinfo{author}{\bibfnamefont{R.~J.} \bibnamefont{Protheroe}},
  \bibnamefont{and} \bibinfo{author}{\bibfnamefont{J.~P.}
  \bibnamefont{Rachen}}, \bibinfo{journal}{Phys.Rev.}
  \textbf{\bibinfo{volume}{D63}}, \bibinfo{pages}{023003}
  (\bibinfo{year}{2000}), \eprint{astro-ph/9812398}.

\bibitem[{\citenamefont{Senno et~al.}(2015{\natexlab{a}})\citenamefont{Senno,
  M\'esz\'aros, Murase, Baerwald, and Rees}}]{Senno:2015tra}
\bibinfo{author}{\bibfnamefont{N.}~\bibnamefont{Senno}},
  \bibinfo{author}{\bibfnamefont{P.}~\bibnamefont{M\'esz\'aros}},
  \bibinfo{author}{\bibfnamefont{K.}~\bibnamefont{Murase}},
  \bibinfo{author}{\bibfnamefont{P.}~\bibnamefont{Baerwald}}, \bibnamefont{and}
  \bibinfo{author}{\bibfnamefont{M.~J.} \bibnamefont{Rees}},
  \bibinfo{journal}{Astrophys.J.} \textbf{\bibinfo{volume}{806}},
  \bibinfo{pages}{24} (\bibinfo{year}{2015}{\natexlab{a}}),
  \eprint{1501.04934}.

\bibitem[{\citenamefont{Costamante}(2013)}]{Costamante:2013sva}
\bibinfo{author}{\bibfnamefont{L.}~\bibnamefont{Costamante}},
  \bibinfo{journal}{Int.J.Mod.Phys.} \textbf{\bibinfo{volume}{D22}},
  \bibinfo{pages}{1330025} (\bibinfo{year}{2013}), \eprint{1309.0612}.

\bibitem[{\citenamefont{Inoue}(2014)}]{Inoue:2014ona}
\bibinfo{author}{\bibfnamefont{Y.}~\bibnamefont{Inoue}} (\bibinfo{year}{2014}),
  \eprint{1412.3886}.

\bibitem[{\citenamefont{Atoyan and Dermer}(2001)}]{Atoyan:2001ey}
\bibinfo{author}{\bibfnamefont{A.}~\bibnamefont{Atoyan}} \bibnamefont{and}
  \bibinfo{author}{\bibfnamefont{C.~D.} \bibnamefont{Dermer}},
  \bibinfo{journal}{Phys.Rev.Lett.} \textbf{\bibinfo{volume}{87}},
  \bibinfo{pages}{221102} (\bibinfo{year}{2001}), \eprint{astro-ph/0108053}.

\bibitem[{\citenamefont{Murase et~al.}(2014)\citenamefont{Murase, Inoue, and
  Dermer}}]{Murase:2014foa}
\bibinfo{author}{\bibfnamefont{K.}~\bibnamefont{Murase}},
  \bibinfo{author}{\bibfnamefont{Y.}~\bibnamefont{Inoue}}, \bibnamefont{and}
  \bibinfo{author}{\bibfnamefont{C.~D.} \bibnamefont{Dermer}},
  \bibinfo{journal}{Phys.Rev.} \textbf{\bibinfo{volume}{D90}},
  \bibinfo{pages}{023007} (\bibinfo{year}{2014}), \eprint{1403.4089}.

\bibitem[{\citenamefont{Di~Mauro et~al.}(2014)\citenamefont{Di~Mauro, Cuoco,
  Donato, and Siegal-Gaskins}}]{DiMauro:2014wha}
\bibinfo{author}{\bibfnamefont{M.}~\bibnamefont{Di~Mauro}},
  \bibinfo{author}{\bibfnamefont{A.}~\bibnamefont{Cuoco}},
  \bibinfo{author}{\bibfnamefont{F.}~\bibnamefont{Donato}}, \bibnamefont{and}
  \bibinfo{author}{\bibfnamefont{J.~M.} \bibnamefont{Siegal-Gaskins}},
  \bibinfo{journal}{J. Cosmol. Astropart. Phys.}
  \textbf{\bibinfo{volume}{1411}}, \bibinfo{pages}{021} (\bibinfo{year}{2014}),
  \eprint{1407.3275}.

\bibitem[{\citenamefont{Ackermann
  et~al.}(2015{\natexlab{b}})}]{Ackermann:2015gga}
\bibinfo{author}{\bibfnamefont{M.}~\bibnamefont{Ackermann}}
  \bibnamefont{et~al.} (\bibinfo{collaboration}{Fermi LAT Collaboration}),
  \bibinfo{journal}{J. Cosmol. Astropart. Phys.}
  \textbf{\bibinfo{volume}{1509}}, \bibinfo{pages}{008}
  (\bibinfo{year}{2015}{\natexlab{b}}), \eprint{1501.05464}.

\bibitem[{\citenamefont{Ajello et~al.}(2015)}]{Ajello:2015mfa}
\bibinfo{author}{\bibfnamefont{M.}~\bibnamefont{Ajello}} \bibnamefont{et~al.},
  \bibinfo{journal}{Astrophys. J.} \textbf{\bibinfo{volume}{800}},
  \bibinfo{pages}{L27} (\bibinfo{year}{2015}), \eprint{1501.05301}.

\bibitem[{\citenamefont{Gaisser}(Cambridge University Press, Cambridge, England
  1990)}]{Gaisser:1990vg}
\bibinfo{author}{\bibfnamefont{T.}~\bibnamefont{Gaisser}},
  \bibinfo{journal}{{\it Cosmic Rays and Particle Physics}}
  (\bibinfo{year}{Cambridge University Press, Cambridge, England 1990}).

\bibitem[{\citenamefont{Dermer and Menon}(Princeton University Press,
  Princeton, NJ, 2009)}]{Dermer:2010hy}
\bibinfo{author}{\bibfnamefont{C.~D.} \bibnamefont{Dermer}} \bibnamefont{and}
  \bibinfo{author}{\bibfnamefont{G.}~\bibnamefont{Menon}},
  \bibinfo{journal}{{\it High Energy Radiation from Black Holes: Gamma Rays,
  Cosmic Rays, and Neutrinos}}  (\bibinfo{year}{Princeton University Press,
  Princeton, NJ, 2009}).

\bibitem[{\citenamefont{Murase et~al.}(2008{\natexlab{b}})\citenamefont{Murase,
  Ioka, Nagataki, and Nakamura}}]{Murase:2008mr}
\bibinfo{author}{\bibfnamefont{K.}~\bibnamefont{Murase}},
  \bibinfo{author}{\bibfnamefont{K.}~\bibnamefont{Ioka}},
  \bibinfo{author}{\bibfnamefont{S.}~\bibnamefont{Nagataki}}, \bibnamefont{and}
  \bibinfo{author}{\bibfnamefont{T.}~\bibnamefont{Nakamura}},
  \bibinfo{journal}{Phys.Rev.} \textbf{\bibinfo{volume}{D78}},
  \bibinfo{pages}{023005} (\bibinfo{year}{2008}{\natexlab{b}}),
  \eprint{0801.2861}.

\bibitem[{\citenamefont{Dermer et~al.}(2014)\citenamefont{Dermer, Murase, and
  Inoue}}]{Dermer:2014vaa}
\bibinfo{author}{\bibfnamefont{C.~D.} \bibnamefont{Dermer}},
  \bibinfo{author}{\bibfnamefont{K.}~\bibnamefont{Murase}}, \bibnamefont{and}
  \bibinfo{author}{\bibfnamefont{Y.}~\bibnamefont{Inoue}},
  \bibinfo{journal}{JHEAp} \textbf{\bibinfo{volume}{3-4}}, \bibinfo{pages}{29}
  (\bibinfo{year}{2014}), \eprint{1406.2633}.

\bibitem[{\citenamefont{Rees}(1967)}]{Rees:1967}
\bibinfo{author}{\bibfnamefont{M.~J.} \bibnamefont{Rees}},
  \bibinfo{journal}{Mon. Not. Roy. Astron. Soc.}
  \textbf{\bibinfo{volume}{135}}, \bibinfo{pages}{345} (\bibinfo{year}{1967}).

\bibitem[{\citenamefont{Rees and M\'esz\'aros}(1992)}]{Rees:1992ek}
\bibinfo{author}{\bibfnamefont{M.~J.} \bibnamefont{Rees}} \bibnamefont{and}
  \bibinfo{author}{\bibfnamefont{P.}~\bibnamefont{M\'esz\'aros}},
  \bibinfo{journal}{Mon. Not. Roy. Astron. Soc.}
  \textbf{\bibinfo{volume}{258}}, \bibinfo{pages}{41} (\bibinfo{year}{1992}).

\bibitem[{\citenamefont{Svensson}(1987)}]{sve87}
\bibinfo{author}{\bibfnamefont{R.}~\bibnamefont{Svensson}},
  \bibinfo{journal}{Mon.Not.Roy.Astron.Soc.} \textbf{\bibinfo{volume}{227}},
  \bibinfo{pages}{403} (\bibinfo{year}{1987}).

\bibitem[{\citenamefont{Dermer et~al.}(2007)\citenamefont{Dermer, Ramirez-Ruiz,
  and Le}}]{Dermer:2007me}
\bibinfo{author}{\bibfnamefont{C.}~\bibnamefont{Dermer}},
  \bibinfo{author}{\bibfnamefont{E.}~\bibnamefont{Ramirez-Ruiz}},
  \bibnamefont{and} \bibinfo{author}{\bibfnamefont{T.}~\bibnamefont{Le}},
  \bibinfo{journal}{Astrophys.J.} \textbf{\bibinfo{volume}{664}},
  \bibinfo{pages}{L67} (\bibinfo{year}{2007}), \eprint{astro-ph/0703219}.

\bibitem[{\citenamefont{Waxman and Bahcall}(1998)}]{Waxman:1998yy}
\bibinfo{author}{\bibfnamefont{E.}~\bibnamefont{Waxman}} \bibnamefont{and}
  \bibinfo{author}{\bibfnamefont{J.~N.} \bibnamefont{Bahcall}},
  \bibinfo{journal}{Phys.Rev.} \textbf{\bibinfo{volume}{D59}},
  \bibinfo{pages}{023002} (\bibinfo{year}{1998}), \eprint{hep-ph/9807282}.

\bibitem[{\citenamefont{Bahcall and Waxman}(2001)}]{Bahcall:1999yr}
\bibinfo{author}{\bibfnamefont{J.~N.} \bibnamefont{Bahcall}} \bibnamefont{and}
  \bibinfo{author}{\bibfnamefont{E.}~\bibnamefont{Waxman}},
  \bibinfo{journal}{Phys.Rev.} \textbf{\bibinfo{volume}{D64}},
  \bibinfo{pages}{023002} (\bibinfo{year}{2001}), \eprint{hep-ph/9902383}.

\bibitem[{\citenamefont{Murase and Beacom}(2010)}]{Murase:2010gj}
\bibinfo{author}{\bibfnamefont{K.}~\bibnamefont{Murase}} \bibnamefont{and}
  \bibinfo{author}{\bibfnamefont{J.~F.} \bibnamefont{Beacom}},
  \bibinfo{journal}{Phys.Rev.} \textbf{\bibinfo{volume}{D81}},
  \bibinfo{pages}{123001} (\bibinfo{year}{2010}), \eprint{1003.4959}.

\bibitem[{\citenamefont{Yoshida and Takami}(2014)}]{Yoshida:2014uka}
\bibinfo{author}{\bibfnamefont{S.}~\bibnamefont{Yoshida}} \bibnamefont{and}
  \bibinfo{author}{\bibfnamefont{H.}~\bibnamefont{Takami}},
  \bibinfo{journal}{Phys.Rev.} \textbf{\bibinfo{volume}{D90}},
  \bibinfo{pages}{123012} (\bibinfo{year}{2014}), \eprint{1409.2950}.

\bibitem[{\citenamefont{Gaisser et~al.}(2013)\citenamefont{Gaisser, Stanev, and
  Tilav}}]{Gaisser:2013bla}
\bibinfo{author}{\bibfnamefont{T.~K.} \bibnamefont{Gaisser}},
  \bibinfo{author}{\bibfnamefont{T.}~\bibnamefont{Stanev}}, \bibnamefont{and}
  \bibinfo{author}{\bibfnamefont{S.}~\bibnamefont{Tilav}},
  \bibinfo{journal}{Front. Phys. China} \textbf{\bibinfo{volume}{8}},
  \bibinfo{pages}{748} (\bibinfo{year}{2013}), \eprint{1303.3565}.

\bibitem[{\citenamefont{Apel et~al.}(2013)}]{Apel:2013ura}
\bibinfo{author}{\bibfnamefont{W.~D.} \bibnamefont{Apel}} \bibnamefont{et~al.},
  \bibinfo{journal}{Phys. Rev.} \textbf{\bibinfo{volume}{D87}},
  \bibinfo{pages}{081101} (\bibinfo{year}{2013}), \eprint{1304.7114}.

\bibitem[{\citenamefont{Katz et~al.}(2013)\citenamefont{Katz, Waxman, Thompson,
  and Loeb}}]{Katz:2013ooa}
\bibinfo{author}{\bibfnamefont{B.}~\bibnamefont{Katz}},
  \bibinfo{author}{\bibfnamefont{E.}~\bibnamefont{Waxman}},
  \bibinfo{author}{\bibfnamefont{T.}~\bibnamefont{Thompson}}, \bibnamefont{and}
  \bibinfo{author}{\bibfnamefont{A.}~\bibnamefont{Loeb}}
  (\bibinfo{year}{2013}), \eprint{1311.0287}.

\bibitem[{\citenamefont{Lacki}(2015)}]{Lacki:2013ata}
\bibinfo{author}{\bibfnamefont{B.~C.} \bibnamefont{Lacki}},
  \bibinfo{journal}{Mon.Not.Roy.Astron.Soc.} \textbf{\bibinfo{volume}{448}},
  \bibinfo{pages}{L20} (\bibinfo{year}{2015}), \eprint{1304.6142}.

\bibitem[{\citenamefont{Ueda et~al.}(2014)\citenamefont{Ueda, Akiyama,
  Hasinger, Miyaji, and Watson}}]{Ueda:2014tma}
\bibinfo{author}{\bibfnamefont{Y.}~\bibnamefont{Ueda}},
  \bibinfo{author}{\bibfnamefont{M.}~\bibnamefont{Akiyama}},
  \bibinfo{author}{\bibfnamefont{G.}~\bibnamefont{Hasinger}},
  \bibinfo{author}{\bibfnamefont{T.}~\bibnamefont{Miyaji}}, \bibnamefont{and}
  \bibinfo{author}{\bibfnamefont{M.~G.} \bibnamefont{Watson}},
  \bibinfo{journal}{Astrophys.J.} \textbf{\bibinfo{volume}{786}},
  \bibinfo{pages}{104} (\bibinfo{year}{2014}), \eprint{1402.1836}.

\bibitem[{\citenamefont{Kalashev et~al.}(2015)\citenamefont{Kalashev, Semikoz,
  and Tkachev}}]{Kalashev:2015cma}
\bibinfo{author}{\bibfnamefont{O.}~\bibnamefont{Kalashev}},
  \bibinfo{author}{\bibfnamefont{D.}~\bibnamefont{Semikoz}}, \bibnamefont{and}
  \bibinfo{author}{\bibfnamefont{I.}~\bibnamefont{Tkachev}},
  \bibinfo{journal}{J.Exp.Theor.Phys.} \textbf{\bibinfo{volume}{120}},
  \bibinfo{pages}{541} (\bibinfo{year}{2015}).

\bibitem[{\citenamefont{Tamborra et~al.}(2014)\citenamefont{Tamborra, Ando, and
  Murase}}]{Tamborra:2014xia}
\bibinfo{author}{\bibfnamefont{I.}~\bibnamefont{Tamborra}},
  \bibinfo{author}{\bibfnamefont{S.}~\bibnamefont{Ando}}, \bibnamefont{and}
  \bibinfo{author}{\bibfnamefont{K.}~\bibnamefont{Murase}},
  \bibinfo{journal}{J. Cosmol. Astropart. Phys.}
  \textbf{\bibinfo{volume}{1409}}, \bibinfo{pages}{043} (\bibinfo{year}{2014}),
  \eprint{1404.1189}.

\bibitem[{\citenamefont{M\'esz\'aros et~al.}(2015)\citenamefont{M\'esz\'aros,
  Asano, Murase, Fox, Gao, and Senno}}]{Meszaros:2015zka}
\bibinfo{author}{\bibfnamefont{P.}~\bibnamefont{M\'esz\'aros}},
  \bibinfo{author}{\bibfnamefont{K.}~\bibnamefont{Asano}},
  \bibinfo{author}{\bibfnamefont{K.}~\bibnamefont{Murase}},
  \bibinfo{author}{\bibfnamefont{D.}~\bibnamefont{Fox}},
  \bibinfo{author}{\bibfnamefont{H.}~\bibnamefont{Gao}}, \bibnamefont{and}
  \bibinfo{author}{\bibfnamefont{N.}~\bibnamefont{Senno}}
  (\bibinfo{year}{2015}), \eprint{1506.02707}.

\bibitem[{\citenamefont{Waxman}(1995)}]{Waxman:1995vg}
\bibinfo{author}{\bibfnamefont{E.}~\bibnamefont{Waxman}},
  \bibinfo{journal}{Phys.Rev.Lett.} \textbf{\bibinfo{volume}{75}},
  \bibinfo{pages}{386} (\bibinfo{year}{1995}), \eprint{astro-ph/9505082}.

\bibitem[{\citenamefont{Vietri}(1995)}]{Vietri:1995hs}
\bibinfo{author}{\bibfnamefont{M.}~\bibnamefont{Vietri}},
  \bibinfo{journal}{Astrophys. J.} \textbf{\bibinfo{volume}{453}},
  \bibinfo{pages}{883} (\bibinfo{year}{1995}), \eprint{astro-ph/9506081}.

\bibitem[{\citenamefont{Aartsen et~al.}(2015{\natexlab{e}})}]{Aartsen:2014aqy}
\bibinfo{author}{\bibfnamefont{M.}~\bibnamefont{Aartsen}} \bibnamefont{et~al.}
  (\bibinfo{collaboration}{IceCube Collaboration}),
  \bibinfo{journal}{Astrophys.J.} \textbf{\bibinfo{volume}{805}},
  \bibinfo{pages}{L5} (\bibinfo{year}{2015}{\natexlab{e}}), \eprint{1412.6510}.

\bibitem[{\citenamefont{Murase et~al.}(2006)\citenamefont{Murase, Ioka,
  Nagataki, and Nakamura}}]{Murase:2006mm}
\bibinfo{author}{\bibfnamefont{K.}~\bibnamefont{Murase}},
  \bibinfo{author}{\bibfnamefont{K.}~\bibnamefont{Ioka}},
  \bibinfo{author}{\bibfnamefont{S.}~\bibnamefont{Nagataki}}, \bibnamefont{and}
  \bibinfo{author}{\bibfnamefont{T.}~\bibnamefont{Nakamura}},
  \bibinfo{journal}{Astrophys.J.} \textbf{\bibinfo{volume}{651}},
  \bibinfo{pages}{L5} (\bibinfo{year}{2006}), \eprint{astro-ph/0607104}.

\bibitem[{\citenamefont{Gupta and Zhang}(2007)}]{Gupta:2006jm}
\bibinfo{author}{\bibfnamefont{N.}~\bibnamefont{Gupta}} \bibnamefont{and}
  \bibinfo{author}{\bibfnamefont{B.}~\bibnamefont{Zhang}},
  \bibinfo{journal}{Astropart.Phys.} \textbf{\bibinfo{volume}{27}},
  \bibinfo{pages}{386} (\bibinfo{year}{2007}), \eprint{astro-ph/0606744}.

\bibitem[{\citenamefont{Bhattacharya et~al.}(2015)\citenamefont{Bhattacharya,
  Enberg, Reno, and Sarcevic}}]{Bhattacharya:2014sta}
\bibinfo{author}{\bibfnamefont{A.}~\bibnamefont{Bhattacharya}},
  \bibinfo{author}{\bibfnamefont{R.}~\bibnamefont{Enberg}},
  \bibinfo{author}{\bibfnamefont{M.~H.} \bibnamefont{Reno}}, \bibnamefont{and}
  \bibinfo{author}{\bibfnamefont{I.}~\bibnamefont{Sarcevic}},
  \bibinfo{journal}{J. Cosmol. Astropart. Phys.}
  \textbf{\bibinfo{volume}{1506}}, \bibinfo{pages}{034} (\bibinfo{year}{2015}),
  \eprint{1407.2985}.

\bibitem[{\citenamefont{Nakar}(2015)}]{Nakar:2015tma}
\bibinfo{author}{\bibfnamefont{E.}~\bibnamefont{Nakar}},
  \bibinfo{journal}{Astrophys. J.} \textbf{\bibinfo{volume}{807}},
  \bibinfo{pages}{172} (\bibinfo{year}{2015}), \eprint{1503.00441}.

\bibitem[{\citenamefont{Senno et~al.}(2015{\natexlab{b}})\citenamefont{Senno,
  Murase, and Meszaros}}]{Senno:2015tsn}
\bibinfo{author}{\bibfnamefont{N.}~\bibnamefont{Senno}},
  \bibinfo{author}{\bibfnamefont{K.}~\bibnamefont{Murase}}, \bibnamefont{and}
  \bibinfo{author}{\bibfnamefont{P.}~\bibnamefont{Meszaros}}
  (\bibinfo{year}{2015}{\natexlab{b}}), \eprint{1512.08513}.

\bibitem[{\citenamefont{Murase et~al.}(2009)\citenamefont{Murase, M\'esz\'aros,
  and Zhang}}]{Murase:2009pg}
\bibinfo{author}{\bibfnamefont{K.}~\bibnamefont{Murase}},
  \bibinfo{author}{\bibfnamefont{P.}~\bibnamefont{M\'esz\'aros}},
  \bibnamefont{and} \bibinfo{author}{\bibfnamefont{B.}~\bibnamefont{Zhang}},
  \bibinfo{journal}{Phys.Rev.} \textbf{\bibinfo{volume}{D79}},
  \bibinfo{pages}{103001} (\bibinfo{year}{2009}), \eprint{0904.2509}.

\bibitem[{\citenamefont{Fang et~al.}(2014)\citenamefont{Fang, Kotera, Murase,
  and Olinto}}]{Fang:2013vla}
\bibinfo{author}{\bibfnamefont{K.}~\bibnamefont{Fang}},
  \bibinfo{author}{\bibfnamefont{K.}~\bibnamefont{Kotera}},
  \bibinfo{author}{\bibfnamefont{K.}~\bibnamefont{Murase}}, \bibnamefont{and}
  \bibinfo{author}{\bibfnamefont{A.~V.} \bibnamefont{Olinto}},
  \bibinfo{journal}{Phys.Rev.} \textbf{\bibinfo{volume}{D90}},
  \bibinfo{pages}{103005} (\bibinfo{year}{2014}), \eprint{1311.2044}.

\bibitem[{\citenamefont{Liang et~al.}(2007)\citenamefont{Liang, Zhang, and
  Dai}}]{Liang:2006ci}
\bibinfo{author}{\bibfnamefont{E.}~\bibnamefont{Liang}},
  \bibinfo{author}{\bibfnamefont{B.}~\bibnamefont{Zhang}}, \bibnamefont{and}
  \bibinfo{author}{\bibfnamefont{Z.}~\bibnamefont{Dai}},
  \bibinfo{journal}{Astrophys.J.} \textbf{\bibinfo{volume}{662}},
  \bibinfo{pages}{1111} (\bibinfo{year}{2007}), \eprint{astro-ph/0605200}.

\bibitem[{\citenamefont{Smith et~al.}(2011)\citenamefont{Smith, Li, Filippenko,
  and Chornock}}]{Smith:2010vz}
\bibinfo{author}{\bibfnamefont{N.}~\bibnamefont{Smith}},
  \bibinfo{author}{\bibfnamefont{W.}~\bibnamefont{Li}},
  \bibinfo{author}{\bibfnamefont{A.~V.} \bibnamefont{Filippenko}},
  \bibnamefont{and} \bibinfo{author}{\bibfnamefont{R.}~\bibnamefont{Chornock}},
  \bibinfo{journal}{Mon.Not.Roy.Astron.Soc.} \textbf{\bibinfo{volume}{412}},
  \bibinfo{pages}{1522} (\bibinfo{year}{2011}), \eprint{1006.3899}.

\bibitem[{\citenamefont{Matsumoto et~al.}(2015)\citenamefont{Matsumoto,
  Nakauchi, Ioka, Heger, and Nakamura}}]{Matsumoto:2015bga}
\bibinfo{author}{\bibfnamefont{T.}~\bibnamefont{Matsumoto}},
  \bibinfo{author}{\bibfnamefont{D.}~\bibnamefont{Nakauchi}},
  \bibinfo{author}{\bibfnamefont{K.}~\bibnamefont{Ioka}},
  \bibinfo{author}{\bibfnamefont{A.}~\bibnamefont{Heger}}, \bibnamefont{and}
  \bibinfo{author}{\bibfnamefont{T.}~\bibnamefont{Nakamura}},
  \bibinfo{journal}{Astrophys. J.} \textbf{\bibinfo{volume}{810}},
  \bibinfo{pages}{64} (\bibinfo{year}{2015}), \eprint{1506.05802}.

\bibitem[{\citenamefont{Iocco et~al.}(2008)\citenamefont{Iocco, Murase,
  Nagataki, and Serpico}}]{Iocco:2007td}
\bibinfo{author}{\bibfnamefont{F.}~\bibnamefont{Iocco}},
  \bibinfo{author}{\bibfnamefont{K.}~\bibnamefont{Murase}},
  \bibinfo{author}{\bibfnamefont{S.}~\bibnamefont{Nagataki}}, \bibnamefont{and}
  \bibinfo{author}{\bibfnamefont{P.~D.} \bibnamefont{Serpico}},
  \bibinfo{journal}{Astrophys. J.} \textbf{\bibinfo{volume}{675}},
  \bibinfo{pages}{937} (\bibinfo{year}{2008}), \eprint{0707.0515}.

\bibitem[{\citenamefont{Ishihara}(2015)}]{Ishihara:2015tev}
\bibinfo{author}{\bibfnamefont{A.}~\bibnamefont{Ishihara}}
  (\bibinfo{collaboration}{IceCube Collaboration}), \bibinfo{journal}{in talks
  presented at the TeV Particle Astrophysics 2015}  (\bibinfo{year}{2015}).

\bibitem[{\citenamefont{Murase}(2015{\natexlab{b}})}]{Murase:2015ndr}
\bibinfo{author}{\bibfnamefont{K.}~\bibnamefont{Murase}}
  (\bibinfo{year}{2015}{\natexlab{b}}), \eprint{1511.01590}.

\bibitem[{\citenamefont{Inoue}(2008)}]{Inoue:2008ks}
\bibinfo{author}{\bibfnamefont{S.}~\bibnamefont{Inoue}},
  \bibinfo{journal}{J.Phys.Conf.Ser.} \textbf{\bibinfo{volume}{120}},
  \bibinfo{pages}{062001} (\bibinfo{year}{2008}), \eprint{0809.3205}.

\bibitem[{\citenamefont{Ajello et~al.}(2014)\citenamefont{Ajello, Romani,
  Gasparrini, Shaw, Bolmer et~al.}}]{Ajello:2013lka}
\bibinfo{author}{\bibfnamefont{M.}~\bibnamefont{Ajello}},
  \bibinfo{author}{\bibfnamefont{R.}~\bibnamefont{Romani}},
  \bibinfo{author}{\bibfnamefont{D.}~\bibnamefont{Gasparrini}},
  \bibinfo{author}{\bibfnamefont{M.}~\bibnamefont{Shaw}},
  \bibinfo{author}{\bibfnamefont{J.}~\bibnamefont{Bolmer}},
  \bibnamefont{et~al.}, \bibinfo{journal}{Astrophys.J.}
  \textbf{\bibinfo{volume}{780}}, \bibinfo{pages}{73} (\bibinfo{year}{2014}),
  \eprint{1310.0006}.

\bibitem[{\citenamefont{Tavecchio and Ghisellini}(2015)}]{Tavecchio:2014eia}
\bibinfo{author}{\bibfnamefont{F.}~\bibnamefont{Tavecchio}} \bibnamefont{and}
  \bibinfo{author}{\bibfnamefont{G.}~\bibnamefont{Ghisellini}},
  \bibinfo{journal}{Mon.Not.Roy.Astron.Soc.} \textbf{\bibinfo{volume}{451}},
  \bibinfo{pages}{1502} (\bibinfo{year}{2015}), \eprint{1411.2783}.

\bibitem[{\citenamefont{Blandford and Konigl}(1979)}]{Blandford:1979za}
\bibinfo{author}{\bibfnamefont{R.~D.} \bibnamefont{Blandford}}
  \bibnamefont{and} \bibinfo{author}{\bibfnamefont{A.}~\bibnamefont{Konigl}},
  \bibinfo{journal}{Astrophys. J.} \textbf{\bibinfo{volume}{232}},
  \bibinfo{pages}{34} (\bibinfo{year}{1979}).

\bibitem[{\citenamefont{Padovani et~al.}(2015)\citenamefont{Padovani,
  Petropoulou, Giommi, and Resconi}}]{Padovani:2015mba}
\bibinfo{author}{\bibfnamefont{P.}~\bibnamefont{Padovani}},
  \bibinfo{author}{\bibfnamefont{M.}~\bibnamefont{Petropoulou}},
  \bibinfo{author}{\bibfnamefont{P.}~\bibnamefont{Giommi}}, \bibnamefont{and}
  \bibinfo{author}{\bibfnamefont{E.}~\bibnamefont{Resconi}},
  \textbf{\bibinfo{volume}{452}}, \bibinfo{pages}{1877} (\bibinfo{year}{2015}),
  \eprint{1506.09135}.

\bibitem[{\citenamefont{Tavecchio et~al.}(2014)\citenamefont{Tavecchio,
  Ghisellini, and Guetta}}]{Tavecchio:2014xha}
\bibinfo{author}{\bibfnamefont{F.}~\bibnamefont{Tavecchio}},
  \bibinfo{author}{\bibfnamefont{G.}~\bibnamefont{Ghisellini}},
  \bibnamefont{and} \bibinfo{author}{\bibfnamefont{D.}~\bibnamefont{Guetta}},
  \bibinfo{journal}{Astrophys.J.} \textbf{\bibinfo{volume}{793}},
  \bibinfo{pages}{L18} (\bibinfo{year}{2014}).

\bibitem[{\citenamefont{Glusenkamp}(2015)}]{Glusenkamp:2015jca}
\bibinfo{author}{\bibfnamefont{T.}~\bibnamefont{Glusenkamp}}
  (\bibinfo{collaboration}{IceCube Collaboration}) (\bibinfo{year}{2015}),
  \eprint{1502.03104}.

\bibitem[{\citenamefont{Stecker et~al.}(1991)\citenamefont{Stecker, Done,
  Salamon, and Sommers}}]{Stecker:1991vm}
\bibinfo{author}{\bibfnamefont{F.~W.} \bibnamefont{Stecker}},
  \bibinfo{author}{\bibfnamefont{C.}~\bibnamefont{Done}},
  \bibinfo{author}{\bibfnamefont{M.~H.} \bibnamefont{Salamon}},
  \bibnamefont{and} \bibinfo{author}{\bibfnamefont{P.}~\bibnamefont{Sommers}},
  \bibinfo{journal}{Phys.Rev.Lett.} \textbf{\bibinfo{volume}{66}},
  \bibinfo{pages}{2697} (\bibinfo{year}{1991}).

\bibitem[{\citenamefont{Alvarez-Muniz and
  M\'esz\'aros}(2004)}]{AlvarezMuniz:2004uz}
\bibinfo{author}{\bibfnamefont{J.}~\bibnamefont{Alvarez-Muniz}}
  \bibnamefont{and}
  \bibinfo{author}{\bibfnamefont{P.}~\bibnamefont{M\'esz\'aros}},
  \bibinfo{journal}{Phys.Rev.} \textbf{\bibinfo{volume}{D70}},
  \bibinfo{pages}{123001} (\bibinfo{year}{2004}), \eprint{astro-ph/0409034}.

\bibitem[{\citenamefont{Pe'er et~al.}(2009)\citenamefont{Pe'er, Murase, and
  M\'esz\'aros}}]{Pe'er:2009rc}
\bibinfo{author}{\bibfnamefont{A.}~\bibnamefont{Pe'er}},
  \bibinfo{author}{\bibfnamefont{K.}~\bibnamefont{Murase}}, \bibnamefont{and}
  \bibinfo{author}{\bibfnamefont{P.}~\bibnamefont{M\'esz\'aros}},
  \bibinfo{journal}{Phys.Rev.} \textbf{\bibinfo{volume}{D80}},
  \bibinfo{pages}{123018} (\bibinfo{year}{2009}), \eprint{0911.1776}.

\bibitem[{\citenamefont{Burrows et~al.}(2011)}]{Burrows:2011dn}
\bibinfo{author}{\bibfnamefont{D.~N.} \bibnamefont{Burrows}}
  \bibnamefont{et~al.}, \bibinfo{journal}{Nature (London)}
  \textbf{\bibinfo{volume}{476}}, \bibinfo{pages}{421} (\bibinfo{year}{2011}),
  \eprint{1104.4787}.

\bibitem[{\citenamefont{Aleksic et~al.}(2014)\citenamefont{Aleksic, Ansoldi,
  Antonelli, Antoranz, Babic et~al.}}]{Aleksic:2014xsg}
\bibinfo{author}{\bibfnamefont{J.}~\bibnamefont{Aleksic}},
  \bibinfo{author}{\bibfnamefont{S.}~\bibnamefont{Ansoldi}},
  \bibinfo{author}{\bibfnamefont{L.}~\bibnamefont{Antonelli}},
  \bibinfo{author}{\bibfnamefont{P.}~\bibnamefont{Antoranz}},
  \bibinfo{author}{\bibfnamefont{A.}~\bibnamefont{Babic}},
  \bibnamefont{et~al.}, \bibinfo{journal}{Science}
  \textbf{\bibinfo{volume}{346}}, \bibinfo{pages}{1080} (\bibinfo{year}{2014}),
  \eprint{1412.4936}.

\bibitem[{\citenamefont{Schatz et~al.}(Universal Academy Press,
  2003)\citenamefont{Schatz, Fessler, Antoni, Apel, Badea
  et~al.}}]{Schatz:2003aw}
\bibinfo{author}{\bibfnamefont{G.}~\bibnamefont{Schatz}},
  \bibinfo{author}{\bibfnamefont{F.}~\bibnamefont{Fessler}},
  \bibinfo{author}{\bibfnamefont{T.}~\bibnamefont{Antoni}},
  \bibinfo{author}{\bibfnamefont{W.}~\bibnamefont{Apel}},
  \bibinfo{author}{\bibfnamefont{F.}~\bibnamefont{Badea}}, \bibnamefont{et~al.}
  (\bibinfo{collaboration}{KASCADE Collaboration}), \bibinfo{journal}{{\it
  Proceedings of ICRC 2003, Frontiers Science Series, Tokyo, Japan}} p.
  \bibinfo{pages}{2293} (\bibinfo{year}{Universal Academy Press, 2003}).

\bibitem[{\citenamefont{Chantell et~al.}(1997)\citenamefont{Chantell, Covault,
  Cronin, Fick, Fortson et~al.}}]{Chantell:1997gs}
\bibinfo{author}{\bibfnamefont{M.~C.} \bibnamefont{Chantell}},
  \bibinfo{author}{\bibfnamefont{C.~E.} \bibnamefont{Covault}},
  \bibinfo{author}{\bibfnamefont{J.~W.} \bibnamefont{Cronin}},
  \bibinfo{author}{\bibfnamefont{B.~E.} \bibnamefont{Fick}},
  \bibinfo{author}{\bibfnamefont{L.~F.} \bibnamefont{Fortson}},
  \bibnamefont{et~al.}, \bibinfo{journal}{Phys. Rev. Lett.}
  \textbf{\bibinfo{volume}{79}}, \bibinfo{pages}{1805} (\bibinfo{year}{1997}),
  \eprint{astro-ph/9705246}.

\bibitem[{\citenamefont{Borione et~al.}(1998)\citenamefont{Borione, Catanese,
  Chantell, Covault, Cronin et~al.}}]{Borione:1997fy}
\bibinfo{author}{\bibfnamefont{A.}~\bibnamefont{Borione}},
  \bibinfo{author}{\bibfnamefont{M.}~\bibnamefont{Catanese}},
  \bibinfo{author}{\bibfnamefont{M.}~\bibnamefont{Chantell}},
  \bibinfo{author}{\bibfnamefont{C.}~\bibnamefont{Covault}},
  \bibinfo{author}{\bibfnamefont{J.}~\bibnamefont{Cronin}},
  \bibnamefont{et~al.}, \bibinfo{journal}{Astrophys.J.}
  \textbf{\bibinfo{volume}{493}}, \bibinfo{pages}{175} (\bibinfo{year}{1998}),
  \eprint{astro-ph/9703063}.

\bibitem[{\citenamefont{Ahlers et~al.}(2016)\citenamefont{Ahlers, Bai, Barger,
  and Lu}}]{Ahlers:2015moa}
\bibinfo{author}{\bibfnamefont{M.}~\bibnamefont{Ahlers}},
  \bibinfo{author}{\bibfnamefont{Y.}~\bibnamefont{Bai}},
  \bibinfo{author}{\bibfnamefont{V.}~\bibnamefont{Barger}}, \bibnamefont{and}
  \bibinfo{author}{\bibfnamefont{R.}~\bibnamefont{Lu}}, \bibinfo{journal}{Phys.
  Rev.} \textbf{\bibinfo{volume}{D93}}, \bibinfo{pages}{013009}
  (\bibinfo{year}{2016}), \eprint{1505.03156}.

\bibitem[{\citenamefont{Chen et~al.}(2015)\citenamefont{Chen, Bhupal~Dev, and
  Soni}}]{Chen:2014gxa}
\bibinfo{author}{\bibfnamefont{C.-Y.} \bibnamefont{Chen}},
  \bibinfo{author}{\bibfnamefont{P.~S.} \bibnamefont{Bhupal~Dev}},
  \bibnamefont{and} \bibinfo{author}{\bibfnamefont{A.}~\bibnamefont{Soni}},
  \bibinfo{journal}{Phys. Rev.} \textbf{\bibinfo{volume}{D92}},
  \bibinfo{pages}{073001} (\bibinfo{year}{2015}), \eprint{1411.5658}.

\bibitem[{\citenamefont{Aartsen et~al.}(2014{\natexlab{b}})}]{Aartsen:2014njl}
\bibinfo{author}{\bibfnamefont{M.}~\bibnamefont{Aartsen}} \bibnamefont{et~al.}
  (\bibinfo{collaboration}{IceCube-Gen2 Collaboration})
  (\bibinfo{year}{2014}{\natexlab{b}}), \eprint{1412.5106}.

\bibitem[{\citenamefont{Ahlers and Halzen}(2014)}]{Ahlers:2014ioa}
\bibinfo{author}{\bibfnamefont{M.}~\bibnamefont{Ahlers}} \bibnamefont{and}
  \bibinfo{author}{\bibfnamefont{F.}~\bibnamefont{Halzen}},
  \bibinfo{journal}{Phys.Rev.} \textbf{\bibinfo{volume}{D90}},
  \bibinfo{pages}{043005} (\bibinfo{year}{2014}), \eprint{1406.2160}.

\bibitem[{\citenamefont{Murase and Waxman}(to be published)}]{mw15}
\bibinfo{author}{\bibfnamefont{K.}~\bibnamefont{Murase}} \bibnamefont{and}
  \bibinfo{author}{\bibfnamefont{E.}~\bibnamefont{Waxman}} (\bibinfo{year}{to
  be published}).

\bibitem[{\citenamefont{Ando et~al.}(2015)\citenamefont{Ando, Tamborra, and
  Zandanel}}]{Ando:2015bva}
\bibinfo{author}{\bibfnamefont{S.}~\bibnamefont{Ando}},
  \bibinfo{author}{\bibfnamefont{I.}~\bibnamefont{Tamborra}}, \bibnamefont{and}
  \bibinfo{author}{\bibfnamefont{F.}~\bibnamefont{Zandanel}},
  \bibinfo{journal}{Phys. Rev. Lett.} \textbf{\bibinfo{volume}{115}},
  \bibinfo{pages}{221101} (\bibinfo{year}{2015}), \eprint{1509.02444}.

\bibitem[{\citenamefont{Bechtol et~al.}(2015)\citenamefont{Bechtol, Ahlers,
  Di~Mauro, Ajello, and Vandenbroucke}}]{Bechtol:2015uqb}
\bibinfo{author}{\bibfnamefont{K.}~\bibnamefont{Bechtol}},
  \bibinfo{author}{\bibfnamefont{M.}~\bibnamefont{Ahlers}},
  \bibinfo{author}{\bibfnamefont{M.}~\bibnamefont{Di~Mauro}},
  \bibinfo{author}{\bibfnamefont{M.}~\bibnamefont{Ajello}}, \bibnamefont{and}
  \bibinfo{author}{\bibfnamefont{J.}~\bibnamefont{Vandenbroucke}}
  (\bibinfo{year}{2015}), \eprint{1511.00688}.

\bibitem[{\citenamefont{M\'esz\'aros and Waxman}(2001)}]{Meszaros:2001ms}
\bibinfo{author}{\bibfnamefont{P.}~\bibnamefont{M\'esz\'aros}}
  \bibnamefont{and} \bibinfo{author}{\bibfnamefont{E.}~\bibnamefont{Waxman}},
  \bibinfo{journal}{Phys.Rev.Lett.} \textbf{\bibinfo{volume}{87}},
  \bibinfo{pages}{171102} (\bibinfo{year}{2001}), \eprint{astro-ph/0103275}.

\bibitem[{\citenamefont{Murase and Beacom}(2013)}]{Murase:2012rd}
\bibinfo{author}{\bibfnamefont{K.}~\bibnamefont{Murase}} \bibnamefont{and}
  \bibinfo{author}{\bibfnamefont{J.~F.} \bibnamefont{Beacom}},
  \bibinfo{journal}{J. Cosmol. Astropart. Phys.}
  \textbf{\bibinfo{volume}{1302}}, \bibinfo{pages}{028} (\bibinfo{year}{2013}),
  \eprint{1209.0225}.

\bibitem[{\citenamefont{Lacki and Thompson}(2013)}]{Lacki:2010ue}
\bibinfo{author}{\bibfnamefont{B.~C.} \bibnamefont{Lacki}} \bibnamefont{and}
  \bibinfo{author}{\bibfnamefont{T.~A.} \bibnamefont{Thompson}},
  \bibinfo{journal}{Astrophys.J.} \textbf{\bibinfo{volume}{762}},
  \bibinfo{pages}{29} (\bibinfo{year}{2013}), \eprint{1010.3030}.

\end{thebibliography}

\clearpage
\section{Supplementary Material}
All the key points are described in the main text, which are general and not sensitive to details of the astrophysical models.  We here describe possible scenarios for TeV-PeV neutrino sources that are obscured in GeV-TeV $\gamma$ rays, without going through specific details.  For candidate sources of CR reservoirs including starburst galaxies and galaxy clusters, see Refs.~\cite{Murase:2013rfa,Tamborra:2014xia} and references therein.  

\subsection{Candidates of Hidden Cosmic-Ray Accelerators}\label{sec4}
By Equations~(1) and (2) in the main text, the diffuse (all-flavor) neutrino flux from $p\gamma$ sources is estimated to be
\begin{multline}
E_{\nu}^2\Phi_{\nu}\simeq0.76\times{10}^{-7}~{\rm GeV}~{\rm cm}^{-2}~{\rm s}^{-1}~{\rm sr}^{-1}~\\
\times{\rm min}[1,f_{p\gamma}]f_{\rm sup}\left(\frac{\xi_z}{3}\right)\left(\frac{\varepsilon_pQ_{\varepsilon_p}}{{10}^{44}~{\rm erg}~{\rm Mpc}^{-3}~{\rm yr}^{-1}}\right)\,,
\nonumber
\end{multline}
where $f_{\rm sup}(\leq1)$ is the suppression factor due to meson and muon cooling, $\xi_z$ is a factor accounting for redshift evolution of the source density~\cite{Waxman:1998yy,Bahcall:1999yr}. For no redshift evolution, we have $\xi_z\simeq0.6$.  For the star-formation history and flat spectrum radio quasar evolution we obtain $\xi_z\simeq3$ and $\xi_z\simeq8$, respectively.  In the following we will discuss specific scenarios in terms of their CR luminosity density $\varepsilon_{p}Q_{\varepsilon_p}$ and photomeson production efficiency $f_{p\gamma}$.

At present, there are several models that can explain the $10\mbox{--}100$~TeV neutrino data.   
For a power-law proton spectrum, the total CR luminosity density (at $z=0$) is expressed by $Q_p=(\varepsilon_pQ_{\varepsilon_p}){\mathcal R}_{p}$, where ${\mathcal R}_p(\varepsilon_p)$ is the conversion factor; ${\mathcal R}_{p}=\ln(\varepsilon_p^{\rm max}/\varepsilon_p^{\rm min})$ for $s_{\rm cr}=2$ and ${\mathcal R}_{p}={(\varepsilon_p/\varepsilon_p^{\rm min})}^{s_{\rm cr}-2}[1-{(\varepsilon_p^{\rm max}/\varepsilon_p^{\rm min})}^{2-s_{\rm cr}}]/(s_{\rm cr}-2)$ for $s_{\rm cr}>2$.  In the shock acceleration theory, one typically expects $\varepsilon_p^{\rm min}\sim\Gamma m_pc^2$ or $\Gamma^2m_pc^2$.  For example, assuming $\varepsilon_p^{\rm max}=60$~PeV, $s_{\rm cr}=2$ and $\varepsilon_p^{\rm min}=1$~TeV lead to ${\mathcal R}_p\sim10$, while $s_{\rm cr}=2.5$ and $\varepsilon_p^{\rm min}=1$~TeV give ${\mathcal R}_p\sim10$ at 25~TeV.  We hereafter use ${\mathcal R}_p\sim10$ as a fiducial value, although lower $\varepsilon_p^{\rm min}$ (e.g., $\sim1$~GeV) leads to larger ${\mathcal R}_p$. 

{\it Choked jets and newborn pulsars.---}
Massive star explosions such as supernovae and GRBs are considered as promising sites of CR acceleration.  GRB prompt emission is believed to be high-energy radiation from expanding relativistic outflows launched by a black hole with an accretion disk or a fast-rotating magnetar.  Particle acceleration may occur both at internal shocks inside a relativistic outflow and a pair of external shocks caused by the outflow, via the shock acceleration and/or magnetic reconnections~\cite{Meszaros:2015zka}.  
GRBs may explain UHE CRs~\cite{Waxman:1995vg,Vietri:1995hs}, since their integrated $\gamma$-ray luminosity density $Q_{\gamma}\sim{10}^{44}~{\rm erg}~{\rm Mpc}^{-3}~{\rm yr}^{-1}$ is comparable to the differential UHE CR luminosity density $\varepsilon_{\rm cr}Q_{\varepsilon_{\rm cr}}\sim0.5\times{10}^{44}~{\rm erg}~{\rm Mpc}^{-3}~{\rm yr}^{-1}$ at ${10}^{19.5}$~eV.  
However, stacking analyses for observed GRBs lead to stringent constraints.  It was shown that classical GRBs can contribute $\lesssim1$\% of the observed diffuse neutrino flux~\cite{Aartsen:2014aqy}. 

However, these limits do not apply to low-luminosity GRBs and ultralong GRBs.  Low-power GRBs may have different origins, and most of them are missed by GRB satellites such as {\it Fermi} and {\it Swift}.  Their energy budget may be comparable to that of classical GRBs, so it is possible that they have a significant contribution to the diffuse neutrino flux~\cite{Murase:2006mm,Gupta:2006jm}.  Theoretically, lower-power jets are more difficult to penetrate the progenitor, so it is natural to expect ``choked jets''~\cite{Meszaros:2001ms}.  Although too powerful jets lead to radiation-mediated shocks and do not allow efficient CR acceleration, since all protons can be depleted for meson production, choked GRB jets can account for the IceCube data~\cite{Murase:2013ffa,Murase:2014tsa,Bhattacharya:2014sta,Nakar:2015tma,Senno:2015tsn}.  
Not only jets but also newborn pulsar winds can serve as hidden CR accelerators~\cite{Murase:2009pg,Fang:2013vla}.  The pulsar wind with $\Gamma\sim{10}^{6}$ lead to $\sim50$~TeV neutrinos in the presence of nonthermal target photons generated in the nebula. 

\begin{figure}[t]
\includegraphics[width=\linewidth]{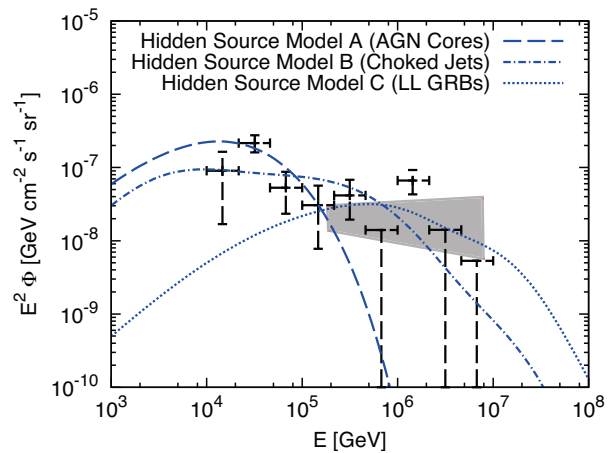}
\caption{All-flavor neutrino fluxes for some $\gamma$-ray obscured $p\gamma$ scenarios that may account for the latest IceCube data~\cite{Aartsen:2015ita}.  The latest data on upgoing neutrinos are also shown~\cite{Ishihara:2015tev}.  We show curves of the AGN core model~\cite{Kimura:2014jba,Murase:2015ndr}, choked jet model~\cite{Murase:2013ffa,Murase:2014tsa}, and low-luminosity GRB model~\cite{Murase:2006mm,Murase:2014tsa}.  Note that model uncertainties are large and contributions to the sub-TeV IGRB are sufficiently small in these models.
\label{fig4}
}
\end{figure}

Such jet-driven and pulsar-driven supernovae have been suggested as origins of low-luminosity GRBs (that are often classified as transrelativistic supernovae) and hypernovae, whose local rates are $\sim{10}^2\mbox{--}{10}^3~{\rm Gpc}^{-3}~{\rm yr}^{-1}$~\cite{Liang:2006ci} and $\sim4000~{\rm Gpc}^{-3}~{\rm yr}^{-1}$~\cite{Smith:2010vz}, respectively.  The available energy budget is $\sim4\times{10}^{46}~{\rm erg}~{\rm Mpc}^{-3}~{\rm yr}^{-1}$, so we expect $\varepsilon_pQ_{\varepsilon_{p}}\lesssim4\times{10}^{45}~{\mathcal R}_{p,1}^{-1}~{\rm erg}~{\rm Mpc}^{-3}~{\rm yr}^{-1}$.  This does not violate the total CR luminosity density of galaxies, $\varepsilon_pQ_{\varepsilon_{p}}\approx{10}^{45}\mbox{--}{10}^{46}~{\rm erg}~{\rm Mpc}^{-3}~{\rm yr}^{-1}$ in the $\sim1\mbox{--}10$~GeV range~\cite{Katz:2013ooa,Lacki:2013ata}, and it is possible for choked jets and pulsars to achieve $E_{\nu}^2\Phi_{\nu}\sim{10}^{-7}~{\rm GeV}~{\rm cm}^{-2}~{\rm s}^{-1}~{\rm sr}^{-1}$ (see Fig.~\ref{fig4}). 

In addition, very massive stars born at high redshifts lead to black holes and could launch jets~(e.g., Ref.~\cite{Matsumoto:2015bga}).  We here point out that choked jets from such high-redshift objects could also give a contribution to the diffuse neutrino flux, as considered in Ref.~\cite{Iocco:2007td}.  

{\it Vicinity of supermassive black holes.--}
AGN show broadband nonthermal emission with major contributions to the x-ray and $\gamma$-ray backgrounds.  About 10\% of AGN is thought to have powerful relativistic jets and these radio-load galaxies have a typical density of $\sim{10}^{-4}~{\rm Mpc}^{-3}$ and jet luminosity of $\sim{10}^{44}~{\rm erg}~{\rm s}^{-1}$.  Their energy budget is estimated to be $\sim3\times{10}^{46}~{\rm erg}~{\rm Mpc}^{-3}~{\rm yr}^{-1}$~\cite{Inoue:2008ks}, implying $\varepsilon_pQ_{\varepsilon_{p}}\lesssim3\times{10}^{45}~{\rm erg}~{\rm Mpc}^{-3}~{\rm yr}^{-1}~{\mathcal R}_{p,1}^{-1}$.  This is comparable to the $\gamma$-ray luminosity density of BL Lac objects, $ Q_{\gamma}\sim2\times{10}^{45}~{\rm erg}~{\rm Mpc}^{-3}~{\rm yr}^{-1}$~\cite{Ajello:2013lka}.  Note that the total CR luminosity density is larger by ${\mathcal R}_p$, and large CR loading factors of $\sim10\mbox{--}100$ are indeed required to explain the observed neutrino and/or UHE CR fluxes~\cite{Murase:2014foa,Tavecchio:2014eia}.  The popular jet model explains various multiwavelength data of AGN~\cite{Blandford:1979za}, and associated CR acceleration at inner jets, knots, hot spots, and radio bubbles or lobes has been discussed for decades (see Ref.~\cite{Murase:2015ndr} as a recent review). 
In particular, CR acceleration and neutrino emissions from the inner jets, including effects of both internal and external radiation fields, have been studied in detail in light of the IceCube data~\cite{Murase:2014foa,Tavecchio:2014xha,Tavecchio:2014eia,Padovani:2015mba}.     
However, the blazar origin of diffuse neutrinos has already been constrained by point-source searches~\cite{Botner:2015ipa,Ishihara:2015tev,mw15,Glusenkamp:2015jca}.  In addition, one-zone leptonic and leptohadronic models predict very hard spectra and do not explain the $10\mbox{--}100$~TeV neutrino data~\cite{Murase:2014foa,Tavecchio:2014xha,Padovani:2015mba,Murase:2015ndr}. 

The situation may be different at deeper regions in the vicinity of supermassive black holes.  
Obviously, the photomeson production efficiency is higher at inner radii, $f_{p\gamma}\gtrsim0.1$~\cite{Stecker:1991vm,Kimura:2014jba,AlvarezMuniz:2004uz}, and x rays can be supplied by the black hole accretion disk.  The x-ray background originates from the accretion power of supermassive black holes, in particular radio-quiet AGN, and is much higher than the $\gamma$-ray background.  The 2-10~keV x-ray luminosity density of AGN is $Q_{X}\approx2\times{10}^{46}~{\rm erg}~{\rm Mpc}^{-3}~{\rm yr}^{-1}$~\cite{Ueda:2014tma}.  The x rays are thought to originate mostly from thermal electrons in the hot coronae.  Although there is no compelling evidence for CR acceleration in such dense environments, a fraction of the accretion energy could be used for CRs, and radio-quiet AGN and low-luminosity AGN can give $E_{\nu}^2\Phi_{\nu}\sim{10}^{-7}~{\rm GeV}~{\rm cm}^{-2}~{\rm s}^{-1}~{\rm sr}^{-1}$~\cite{Stecker:2013fxa,Kimura:2014jba,Kalashev:2015cma} given that $\varepsilon_pQ_{\varepsilon_{p}}\lesssim Q_p\lesssim Q_{X}$ (see Fig.~\ref{fig4}).  
Source models and CR acceleration mechanisms in the vicinity of black holes are uncertain.  First, relativistic jets may be common in galaxies hosting supermassive black holes, even if they are weaker in radio-quiet and low-luminosity AGN (that may include normal galaxies), and efficient CR acceleration in compact regions in the jet may occur~\cite{Pe'er:2009rc,Burrows:2011dn}.  Second, as considered in Ref.~\cite{Kimura:2014jba}, stochastic and/or shear acceleration as well as magnetic reconnections may occur in radiatively inefficient accretion flows of low-luminosity AGN disks and radio-quiet AGN coronae.  Third, in sufficiently low-luminosity objects starved for plasma, electrostatic acceleration in a potential gap formed in the black hole magnetosphere may also work~\cite{Aleksic:2014xsg}. 
Regarding the luminosity density of x rays as an upper limit of nonthermal photon outputs from AGN, $\varepsilon_pQ_{\varepsilon_p}\lesssim2\times{10}^{46}~{\rm erg}~{\rm Mpc}^{-3}~{\rm yr}^{-1}$ would be considered to be a reasonable assumption.

\subsection{Multimessenger Limits on Galactic Contributions}\label{sec5}
Multimessenger data indicate that the diffuse neutrino flux measured in IceCube largely comes from extragalactic sources.  Galactic neutrino emission is excepted to be strongly anisotropic except for exotic scenarios like emission from the Galactic halo.  As shown in Ref.~\cite{Murase:2013rfa}, even these scenarios are constrained by the IGRB: the spectral index is required to be $s\lesssim2.0$ if the Galactic emission is quasiisotropic.  In addition, there are upper limits placed by CR air-shower arrays such as KASCADE~\cite{Schatz:2003aw} and CASA-MIA~\cite{Chantell:1997gs} in the TeV-PeV range.  
The isotropic diffuse $\gamma$-ray intensity at $E_\gamma\sim300$~TeV is limited as $E_\gamma^2\Phi_\gamma\lesssim{10}^{-8}~{\rm GeV}~{\rm cm}^{-2}~{\rm s}^{-1}~{\rm sr}^{-1}$.  With $K=2$, an upper limit on the isotropic Galactic halo all-flavor neutrino intensity at $E_\nu\sim150$~TeV is estimated to be $E_\nu^2\Phi_\nu\lesssim1.5\times{10}^{-8}~{\rm GeV}~{\rm cm}^{-2}~{\rm s}^{-1}~{\rm sr}^{-1}$, which is $\lesssim30$\% of the all-flavor neutrino intensity at $E_\nu=100$~TeV for a spectral index of $s=s_{\rm ob}=2.5$.

A significant contribution may also come from the Galactic plane, e.g., by diffuse CRs or unresolved supernova remnants.  CASA-MIA~\cite{Borione:1997fy} gives $E_\gamma^2\Phi_\gamma\lesssim2\times{10}^{-8}~{\rm GeV}~{\rm cm}^{-2}~{\rm s}^{-1}~{\rm sr}^{-1}$ at $E_\gamma\sim200$~TeV for the region $|b|<5^\circ$ and $50^\circ<l<200^\circ$. Assuming the uniform neutrino intensity over the entire Galactic plane $\Delta\Omega$, the Galactic plane neutrino intensity is constrained as $E_\nu^2\Phi_\nu\lesssim2\times{10}^{-9}~{(\Delta\Omega/1~{\rm sr})}~{\rm GeV}~{\rm cm}^{-2}~{\rm s}^{-1}~{\rm sr}^{-1}$ at $E_\nu\sim100$~TeV, which is only $\sim3$\% of the all-flavor neutrino intensity although some special neutrino sources could exist outside the array's field of view.

If we consider a neutrino emission region around the Galactic center or ridge (e.g., Fermi bubbles), the observational CASA-MIA limit is weakened to $E_\gamma^2\Phi_\gamma\lesssim2\times{10}^{-7}~{\rm GeV}~{\rm cm}^{-2}~{\rm s}^{-1}~{\rm sr}^{-1}$ at $E_\gamma\sim300$~TeV, leading to  $E_\nu^2\Phi_\nu\lesssim3\times{10}^{-8}~{(\Delta\Omega/1~{\rm sr})}~{\rm GeV}~{\rm cm}^{-2}~{\rm s}^{-1}~{\rm sr}^{-1}$ at $E_\nu\sim100$~TeV.  Thus, the Galactic center contribution is expected to be $\lesssim40\mbox{--}50$\%.  In this case, a stronger upper limit ($\lesssim25$\%), which is mostly independent of spectral assumptions~\cite{Ahlers:2015moa}, is derived from the distribution of the high-energy starting events~\cite{Aartsen:2013jdh}.

A separate fit of the IceCube data in the Northern and Southern Hemispheres resulted in different best-fit power-law indices with $s_{\rm ob}\sim2.0$ and $s_{\rm ob}\sim2.56$, respectively~\cite{Aartsen:2015ita}.  This could indicate anisotropic emission originating from the Milky Way, in particular the Galactic center or ridge.  However, this asymmetry is not significant at present.  Also, muon neutrino limits~\cite{Ahlers:2015moa} and diffuse $\gamma$-ray limits already indicate that Galactic contributions should be less than $\sim25\mbox{--}50$\%.  

So far, (a) extragalactic CR accelerators such as supernovae, GRBs, and AGN, (b) extragalactic CR reservoirs such as starbursts and galaxy cluster or groups, and (c) Galactic sources, have been considered as possible origins of the IceCube neutrinos.  This work showed that a $\gamma$-ray obscured population of CR accelerators is suggested by the 10-100~TeV neutrino data.  In principle, as hinted from the latest IceCube analyses~\cite{Aartsen:2015ita}, the diffuse neutrino flux may come from multiple source populations~\cite{Murase:2014tsa,Chen:2014gxa}.  Although fine tuning is needed, a possible Galactic component may alleviate the tension in both $pp$ and $p\gamma$ scenarios.  The $\gamma$-ray obscured neutrino source component is relevant at $\lesssim100$~TeV and/or $\gtrsim100$~TeV energies.  If it is dominant only at low energies, the higher-energy hard component may come from CR accelerators or CR reservoirs as discussed in Refs.~\cite{Murase:2014tsa,Murase:2013rfa,mw15}. 

\end{document}